\documentclass[fleqn,usenatbib]{mnras}

\usepackage{newtxtext,newtxmath}

\usepackage[T1]{fontenc}
\usepackage{ae,aecompl}

\usepackage{graphicx}	
\usepackage{amsmath}	
\usepackage{amssymb}	
\usepackage{lscape}
\usepackage{color}
\usepackage[dvipsnames]{xcolor}
\usepackage{array,longtable} 
\usepackage{placeins}
\usepackage{etoolbox}
\makeatletter
\patchcmd\@combinedblfloats{\box\@outputbox}{\unvbox\@outputbox}{}{\errmessage{\noexpand patch failed}}
\makeatother

\usepackage{eso-pic}

\AddToShipoutPictureBG*{%
  \AtPageUpperLeft{%
    \hspace{0.75\paperwidth}%
    \raisebox{-3.5\baselineskip}{%
      \makebox[0pt][l]{\textnormal{DES 2018-0329}}
}}}%

\AddToShipoutPictureBG*{%
  \AtPageUpperLeft{%
    \hspace{0.75\paperwidth}%
    \raisebox{-4.5\baselineskip}{%
      \makebox[0pt][l]{\textnormal{FERMILAB PUB-18-669-AE}}
}}}%

\title[SLSNe from DES]{Superluminous Supernovae from the Dark Energy Survey}

\author[C. R. Angus et. al]{
\parbox{\textwidth}{
\Large
C.~R.~Angus,$^{1}$\thanks{c.r.angus@soton.ac.uk}
M.~Smith,$^{1}$
M.~Sullivan,$^{1}$
C.~Inserra,$^{1}$
P.~Wiseman,$^{1}$
C.~B.~D'Andrea,$^{2}$
B.~P.~Thomas,$^{3}$
R.~C.~Nichol,$^{3}$
L.~Galbany,$^{4}$
M.~Childress,$^{1}$
J.~Asorey,$^{5}$
P.~J.~Brown,$^{6}$
R.~Casas,$^{7,8}$
F.~J.~Castander,$^{7,8}$
C.~Curtin,$^{9}$
C.~Frohmaier,$^{3}$
K.~Glazebrook,$^{9}$
D.~Gruen,$^{10,11}$
C.~Gutierrez,$^{1}$
R.~Kessler,$^{12,13}$
A.~G.~Kim,$^{14}$
C.~Lidman,$^{15}$
E.~Macaulay,$^{3}$
P.~Nugent,$^{14}$
M.~Pursiainen,$^{1}$
M.~Sako,$^{2}$
M.~Soares-Santos,$^{16}$
R.~C.~Thomas,$^{14}$
T.~M.~C.~Abbott,$^{17}$
S.~Avila,$^{3}$
E.~Bertin,$^{18,19}$
D.~Brooks,$^{20}$
E.~Buckley-Geer,$^{21}$
D.~L.~Burke,$^{10,11}$
A.~Carnero~Rosell,$^{22,23}$
J.~Carretero,$^{24}$
L.~N.~da Costa,$^{23,25}$
J.~De~Vicente,$^{22}$
S.~Desai,$^{26}$
H.~T.~Diehl,$^{21}$
P.~Doel,$^{20}$
T.~F.~Eifler,$^{27,28}$
B.~Flaugher,$^{21}$
P.~Fosalba,$^{7,8}$
J.~Frieman,$^{21,13}$
J.~Garc\'ia-Bellido,$^{29}$
R.~A.~Gruendl,$^{30,31}$
J.~Gschwend,$^{23,25}$
W.~G.~Hartley,$^{20,32}$
D.~L.~Hollowood,$^{33}$
K.~Honscheid,$^{34,35}$
B.~Hoyle,$^{36,37}$
D.~J.~James,$^{38}$
K.~Kuehn,$^{39}$
N.~Kuropatkin,$^{21}$
O.~Lahav,$^{40}$
M.~Lima,$^{41,23}$
M.~A.~G.~Maia,$^{23,25}$
M.~March,$^{2}$
J.~L.~Marshall,$^{6}$
F.~Menanteau,$^{30,31}$
C.~J.~Miller,$^{42,43}$
R.~Miquel,$^{44,24}$
R.~L.~C.~Ogando,$^{23,25}$
A.~A.~Plazas,$^{28}$
A.~K.~Romer,$^{45}$
E.~Sanchez,$^{22}$
R.~Schindler,$^{11}$
M.~Schubnell,$^{43}$
F.~Sobreira,$^{46,23}$
E.~Suchyta,$^{47}$
M.~E.~C.~Swanson,$^{31}$
G.~Tarle,$^{43}$
D.~Thomas,$^{3}$
and D.~L.~Tucker$^{21}$
\begin{center} (DES Collaboration) \end{center}
}
\vspace{0.4cm}
\\
\parbox{\textwidth}{
$^{1}$ School of Physics and Astronomy, University of Southampton,  Southampton, SO17 1BJ, UK\\
$^{2}$ Department of Physics and Astronomy, University of Pennsylvania, Philadelphia, PA 19104, USA\\
$^{3}$ Institute of Cosmology and Gravitation, University of Portsmouth, Portsmouth, PO1 3FX, UK\\
$^{4}$ PITT PACC, Department of Physics and Astronomy, University of Pittsburgh, Pittsburgh, PA 15260, USA\\
$^{5}$ Korea Astronomy and Space Science Institute, Yuseong-gu, Daejeon, 305-348, Korea\\
$^{6}$ George P. and Cynthia Woods Mitchell Institute for Fundamental Physics and Astronomy, and Department of Physics and Astronomy, Texas A\&M University, College Station, TX 77843,  USA\\
$^{7}$ Institut d'Estudis Espacials de Catalunya (IEEC), 08034 Barcelona, Spain\\
$^{8}$ Institute of Space Sciences (ICE, CSIC),  Campus UAB, Carrer de Can Magrans, s/n,  08193 Barcelona, Spain\\
$^{9}$ Centre for Astrophysics \& Supercomputing, Swinburne University of Technology, Victoria 3122, Australia\\
$^{10}$ Kavli Institute for Particle Astrophysics \& Cosmology, P. O. Box 2450, Stanford University, Stanford, CA 94305, USA\\
$^{11}$ SLAC National Accelerator Laboratory, Menlo Park, CA 94025, USA\\
$^{12}$ Department of Astronomy and Astrophysics, University of Chicago, Chicago, IL 60637, USA\\
$^{13}$ Kavli Institute for Cosmological Physics, University of Chicago, Chicago, IL 60637, USA\\
$^{14}$ Lawrence Berkeley National Laboratory, 1 Cyclotron Road, Berkeley, CA 94720, USA\\
$^{15}$ The Research School of Astronomy and Astrophysics, Australian National University, ACT 2601, Australia\\
$^{16}$ Brandeis University, Physics Department, 415 South Street, Waltham MA 02453\\
$^{17}$ Cerro Tololo Inter-American Observatory, National Optical Astronomy Observatory, Casilla 603, La Serena, Chile\\
$^{18}$ CNRS, UMR 7095, Institut d'Astrophysique de Paris, F-75014, Paris, France\\
$^{19}$ Sorbonne Universit\'es, UPMC Univ Paris 06, UMR 7095, Institut d'Astrophysique de Paris, F-75014, Paris, France\\
$^{20}$ Department of Physics \& Astronomy, University College London, Gower Street, London, WC1E 6BT, UK\\
$^{21}$ Fermi National Accelerator Laboratory, P. O. Box 500, Batavia, IL 60510, USA\\
$^{22}$ Centro de Investigaciones Energ\'eticas, Medioambientales y Tecnol\'ogicas (CIEMAT), Madrid, Spain\\
$^{23}$ Laborat\'orio Interinstitucional de e-Astronomia - LIneA, Rua Gal. Jos\'e Cristino 77, Rio de Janeiro, RJ - 20921-400, Brazil\\
$^{24}$ Institut de F\'{\i}sica d'Altes Energies (IFAE), The Barcelona Institute of Science and Technology, Campus UAB, 08193 Bellaterra (Barcelona) Spain\\
$^{25}$ Observat\'orio Nacional, Rua Gal. Jos\'e Cristino 77, Rio de Janeiro, RJ - 20921-400, Brazil\\
$^{26}$ Department of Physics, IIT Hyderabad, Kandi, Telangana 502285, India\\
$^{27}$ Department of Astronomy/Steward Observatory, 933 North Cherry Avenue, Tucson, AZ 85721-0065, USA\\
$^{28}$ Jet Propulsion Laboratory, California Institute of Technology, 4800 Oak Grove Dr., Pasadena, CA 91109, USA\\
$^{29}$ Instituto de Fisica Teorica UAM/CSIC, Universidad Autonoma de Madrid, 28049 Madrid, Spain\\
$^{30}$ Department of Astronomy, University of Illinois at Urbana-Champaign, 1002 W. Green Street, Urbana, IL 61801, USA\\
$^{31}$ National Center for Supercomputing Applications, 1205 West Clark St., Urbana, IL 61801, USA\\
$^{32}$ Department of Physics, ETH Zurich, Wolfgang-Pauli-Strasse 16, CH-8093 Zurich, Switzerland\\
$^{33}$ Santa Cruz Institute for Particle Physics, Santa Cruz, CA 95064, USA\\
$^{34}$ Center for Cosmology and Astro-Particle Physics, The Ohio State University, Columbus, OH 43210, USA\\
$^{35}$ Department of Physics, The Ohio State University, Columbus, OH 43210, USA\\
$^{36}$ Max Planck Institute for Extraterrestrial Physics, Giessenbachstrasse, 85748 Garching, Germany\\
$^{37}$ Universit\"ats-Sternwarte, Fakult\"at f\"ur Physik, Ludwig-Maximilians Universit\"at M\"unchen, Scheinerstr. 1, 81679 M\"unchen, Germany\\
$^{38}$ Harvard-Smithsonian Center for Astrophysics, Cambridge, MA 02138, USA\\
$^{39}$ Australian Astronomical Optics, Macquarie University, North Ryde, NSW 2113, Australia\\
$^{40}$ Department of Physics and Astronomy, University College London, Gower Street,  London WC1E 6BT, UK\\
$^{41}$ Departamento de F\'isica Matem\'atica, Instituto de F\'isica, Universidade de S\~ao Paulo, CP 66318, S\~ao Paulo, SP, 05314-970, Brazil\\
$^{42}$ Department of Astronomy, University of Michigan, Ann Arbor, MI 48109, USA\\
$^{43}$ Department of Physics, University of Michigan, Ann Arbor, MI 48109, USA\\
$^{44}$ Instituci\'o Catalana de Recerca i Estudis Avan\c{c}ats, E-08010 Barcelona, Spain\\
$^{45}$ Department of Physics and Astronomy, Pevensey Building, University of Sussex, Brighton, BN1 9QH, UK\\
$^{46}$ Instituto de F\'isica Gleb Wataghin, Universidade Estadual de Campinas, 13083-859, Campinas, SP, Brazil\\
$^{47}$ Computer Science and Mathematics Division, Oak Ridge National Laboratory, Oak Ridge, TN 37831\\
}
}

\date{Accepted 2019 April 30. Received 2019 April 30; in original form 2018 December 10.}

\pubyear{2019}

\begin{document}
\label{firstpage}
\pagerange{\pageref{firstpage}--\pageref{lastpage}}
\maketitle

\begin{abstract}
We present a sample of 21 hydrogen-free superluminous supernovae (SLSNe-I), and one hydrogen-rich SLSN (SLSN-II) detected during the five-year Dark Energy Survey (DES). These SNe, located in the redshift range $0.220<z<1.998$, represent the largest homogeneously-selected sample of SLSN events at high redshift. We present the observed $g,r,i,z$ light curves for these SNe, which we interpolate using Gaussian Processes. The resulting light curves are analysed to determine the luminosity function of SLSN-I, and their evolutionary timescales. The DES SLSN-I sample significantly broadens the distribution of SLSN-I light curve properties when combined with existing samples from the literature. We fit a magnetar model to our SLSNe, and find that this model alone is unable to replicate the behaviour of many of the bolometric light curves. We search the DES SLSN-I light curves for the presence of initial peaks prior to the main light-curve peak. Using a shock breakout model, our Monte Carlo search finds that 3 of our 14 events with pre-max data display such initial peaks. However, 10 events show no evidence for such peaks, in some cases down to an absolute magnitude of $<-16$, suggesting that such features are not ubiquitous to all SLSN-I events. We also identify a red pre-peak feature within the light curve of one SLSN, which is comparable to that observed within SN2018bsz.
 \end{abstract}




\section{Introduction}

The emergence of deeper, higher cadence transient surveys in recent years has overturned the astrophysical community's view of the variable universe, uncovering previously unidentified classes of transient events. Superluminous supernovae (SLSNe) represent one such group of objects. These events are capable of reaching luminosities more than 10 times brighter than classical type Ia supernovae (SNe Ia), with long-lived optical light curves \citep[e.g.,][see \citealt{Howell2017} for a recent review]{Smith2007,Pastorello2010,Quimby2011,Gal-Yam2012}. Originally recognised as objects peaking with an absolute magnitude of less than $-21$ \citep[][]{Gal-Yam2012}, it is now apparent that SLSNe occupy a wider range of luminosities, with peak luminosities reportedly as faint as $M\sim-20$ and evolutionary timescales spanning more than a factor of five \citep[e.g.][]{Nicholl2015B,Inserra2017,Lunnan2017,DeCia2017}.

SLSNe exhibit spectral diversity, with hydrogen-rich (SLSN-II) and hydrogen-poor (SLSN-I) varieties \citep[and some with hydrogen only at later times, e.g.,][]{Yan2015,Yan2017}. The luminosity and hydrogen signature of SLSNe-II can be explained by interaction between the SN ejecta and surrounding circumstellar material \citep[CSM;][]{Ofek2014, Inserra2016}. However, the mechanism behind SLSNe-I remains unclear: SLSNe-I can produce luminosities in excess of $\simeq$10$^{44}$\,erg\,s$^{-1}$, and radiate total energies of $\simeq$10$^{51}$\,erg, exceeding the energies produced by classical core-collapse SNe by a factor of 100. Such luminosities would require several solar masses of $^{56}$Ni to be synthesised during the explosion. Whilst the production of high $^{56}$Ni masses is physically possible under exotic explosions paradigms such as Pair Instability SNe \citep{Kasen2011}, these models fail to replicate many observable properties of SLSN (e.g. light curve evolution), thus an additional energy input is required to boost their luminosities. Such an engine may take the form of accretion onto a central compact object \citep{Dexter2013}, or the spin down of a newly-formed magnetar \citep{Woosley2010,Kasen2010}. The ability of this latter engine to replicate a large fraction of SLSN-I light curves is encouraging \citep[e.g.][]{Inserra2013,Nicholl2013,Nicholl2017B}, while spectroscopic models of magnetar driven SLSNe-I ejecta are consistent with the observed spectra of SLSNe-I at early times \citep{Dessart2012,Mazzali2016,Dessart2019}, although the mechanism through which energy is transferred from magnetar to the ejecta is still poorly understood. 

The optical pre-maximum spectra of SLSNe-I are blue and rather featureless, with characteristic broad \ion{O}{ii} absorption lines \citep{Quimby2011} as the defining feature, but in the ultraviolet (UV) there are several strong absorption features. The strongest lines have been identified as \ion{Fe}{iii}, \ion{C}{iii}, \ion{C}{ii}, \ion{Mn}{ii}, \ion{Si}{iii} and \ion{Mg}{ii} \citep{Quimby2011,Dessart2012,Howell2013,Mazzali2016,Quimby2018}, with possible variation from event to event \citep[see detailed discussion in][]{Quimby2018}. By 30 rest-frame days after maximum light, SLSNe-I then have a tendency to resemble both normal and broad-lined type Ic SNe at peak \citep{Pastorello2010, Liu2016}. There remains discussion over the presence of various possible sub-groups within this spectral class \citep[e.g.,][]{Gal-Yam2012,Inserra2018,Quimby2018,Nicholl2019}. 

Significant diversity can be observed within samples of published SLSNe-I, both in light curve evolution \citep{Lunnan2017,DeCia2017} and spectroscopic behaviour \citep{Nicholl2015B,Inserra2017,Quimby2018}. A large number of single-object studies have also been published, often highlighting \lq unusual\rq\ features about a particular event. This makes it difficult to identify the properties of a \lq typical\rq\ SLSN-I. For instance, the presence of a precursor bump that precedes the rise of the main light curve peak, observed in the optical light curves of some SLSNe-I \citep{Leloudas2012,Nicholl2015B,Smith2016}, has been suggested to be common to all SLSN-I events \citep{Nicholl2016A}, and may have been previously missed (or fallen below the detection limit) in discovery surveys. At present we can only make tentative estimates of a rate \citep{Nicholl2016A} based upon limited heterogeneous samples, and as such the influence of bumps upon the physical mechanisms of SLSNe-I can only be considered on a transient-by-transient basis. 

The presence of slowly-evolving SLSNe-I also complicates the issue. While samples are currently small in number, a significant fraction of these events appear to show complex fluctuations in their late-time light curves. These may be a consequence of changes in opacity as more highly ionized layers of ejecta from hard magnetar radiation reach the edge of the photosphere \citep{Metzger2014}, or could involve a less uniform ejecta structure \citep{Inserra2017}, although more detailed radiative transfer modelling is required to disentangle the two scenarios. Additional instances of slowly-evolving SLSNe-I whose declines appear to be consistent with the decay rate of $^{56}$Ni have emerged \citep{Lunnan2016}. At present it is unclear whether such events represent a different class of transient with a different mechanism for explosion, or whether they are simply the extremes of a more continuous distribution of luminous transients. 

Despite being intrinsically rare \citep{Quimby2013,McCrum2015,Prajs2016}\footnote{Although searches for high-redshift photometric candidates suggest that the rate may increase at redshifts $z>2$ \citep{Cooke2012,Moriya2018}.}, their extreme optical luminosities means that surveys can detect examples across large search volumes, and there are now many discoveries of SLSNe-I. Larger samples ($\sim20$) of SLSNe-I identified within single surveys such as PanSTARRS \citep{Lunnan2017} and the Palomar Transient Factory \citep[PTF,][]{DeCia2017} have provided a base for a collective analysis of homogeneously-selected SLSNe-I. However, due to the respective depths of these surveys, these samples are limited to either the local Universe \citep[PTF,][]{DeCia2017} or to redshifts of $z\sim1.6$ \citep[PanSTARRS,][]{Lunnan2017}. The Dark Energy Survey (DES), whose SN program provides high cadence, deep optical imaging for identifying transients, presents an unrivalled dataset for identifying and studying SLSNe at higher redshift. Here we present the sample of SLSNe-I identified over the five-year duration of DES. We describe our observations and outline our handling of the data in Section~\ref{sect:obs}, and we overview the various analytical techniques in Section~\ref{sect:mod}. We present the derived characteristics of our sample SNe in Sections~\ref{sect:lc_res} and \ref{sect:bumps_res}. The host galaxies of the sample are presented within Section~\ref{sect:hosts}. We discuss our findings and their implications in Section~\ref{sect:dis}, and finally present our conclusions in Section~\ref{sect:con}. 

Throughout this paper we assume a flat $\Lambda$CDM cosmology and adopt values of $H_{0}=70.0$\,km\,s$^{-1}$\,Mpc$^{-1}$, and $\Omega_\textrm{M}=0.3$.

\section{Observations}
\label{sect:obs}

\subsection{The Dark Energy Survey}
\label{sec:des}

The Dark Energy Survey \citep[DES;][]{DES2005,DESoverview2016} is an optical imaging survey covering 5000\,deg$^{2}$ of the southern sky for the purpose of measuring the dark energy equation of state. Observations are carried out using the Dark Energy Camera \citep[DECam;][]{Flaugher2015A} on the 4-m Blanco Telescope at the Cerro Tololo InterAmerican Observatory (CTIO) in Chile. The DES-SN program \citep[DES-SN;][]{Bernstein2012}, which uses $\approx$ 10 per cent of the total survey time, surveys 10 DECam pointings, imaging 27\,deg$^2$ in $g$, $r$, $i$ and $z$ filters with an approximate 7-day cadence. For a detailed discussion of the DES-SN observing strategy see \citet{Kessler2015,Diehl2018}.

Although the depth of DES provides multi-colour light curves of SLSNe out to high redshift, the 5-month observing season (mid-August to early-February each year) can result in some SLSNe subject to temporal edge effects, particularly at high redshift where time dilation stretches the time-span of the transient's visibility in the observer frame. An additional programme, the \lq Search Using DECam for Superluminous Supernovae\rq\ (SUDSS; PI: Sullivan), was used to supplement the standard DES season with additional epochs. These observations have an approximate 14 day cadence, which extends the total coverage on selected fields to approximately eight months. 

All DES survey images are processed within the DES Data Management system \citep[DESDM; ][]{Sevilla2011,Mohr2012,Desai2012,Morganson2018}. The outputs from this are then subject to difference imaging using the \textsc{DiffImg} pipeline \citep{Kessler2015,Goldstein2015}, a standard pipeline for transient detection in the DES-SN fields which uses deep templates from stacked epochs from previous seasons. The exception are Year 1 (Y1) transients, for which templates are drawn from observations taken within Y2. Transients are then identified using Source Extractor \citep[\textsc{SExtractor}][]{Bertin1996}; see \citet{Papadopoulos2015} for further details. 

\subsection{The DES SLSN sample}
\label{sec:dessample}

The DES-SN programme has successfully identified several SLSN candidates \citep[e.g., see][]{Papadopoulos2015, Smith2016, Pan2017, Smith2018}. SLSNe are not explicitly selected for follow up by their luminosity, but were identified and prioritized for spectroscopic follow up based upon combinations of the following criteria:
\begin{enumerate}
	\item Where the SN rise is visible within the data, the light curve must display a slow ($>25$ day) rise time in the {\textit{observer frame}}.
	\item Where only the SN decline is visible within the data, the light curve must be visible for $>25$ days and must exhibit a slow ($<0.1$\,mag\,day$^{-1}$) rate.
	\item The transient is predominantly blue in colour ($g-r$ \textless 1.0 or $r-i$ \textless 1.0 mag).
    \item The transient must exhibit colour evolution during the decline (to avoid AGN contamination).
	\item The transient resides within a faint ($\Delta m_{\mathrm{host-sn}}>0.7$) or undetected host galaxy within the template DES images \footnote{SLSNe-I have shown a strong preference for faint, low mass host galaxies \citep[e.g.]{Neill2011,Chen2013,Lunnan2014,Leloudas2015,Angus2016,Perley2016,Schulze2016}, although see SN2017egm \citep[][]{Chen2017B,Izzo2018}.}.
\end{enumerate}

These criteria are soft cuts designed designed such that they should encapsulate the photometric properties of the vast majority of literature SLSNe\footnote{with the obvious exception of \lq peculiar\rq~ events such as SN~2017egm in a bright host galaxy.} as observed within any given DES season, whilst excluding obvious contaminants such as SNe-Ia and AGN flares. They also employed for the selection of the photometric SLSN sample from DES (Thomas et al. {\textit{in prep.}}). Potential biases in the selection of this sample are discussed within Section~\ref{selecteffects}.

\subsubsection{Spectra}

We triggered spectroscopic observations for a total of 30 SN candidates whose properties met the above criteria across a variety of telescopes under the follow-up programmes designed to catch any SN whose properties did not fall into classical paradigms (i.e. not typical SNe Ia, SNe Ibc or SNe II). In addition, two other SNe were initially identified as potential SN Ia candidates by the Photometric SN IDentification software \citep[\textsc{psnid};][]{Sako2011} used to prioritize spectroscopic follow-up (D'Andrea et al. {\textit{in prep.}}) and were observed under various SN-Ia programmes prior to classification as SLSNe \citep[see also][]{Pan2017}.

Object classifications were performed using \textsc{superfit} \citep{Howell2005}, implementing the comprehensive spectral template library of \cite{Quimby2018}. We assign the classification of SLSN-I based upon their spectral similarity to other known SLSNe-I within the literature: well described by a hot black body continuum at early times, with broad absorption lines produced by C, O, Si, Mg, and Fe with high velocities ($\sim10000$\,km\,s$^{−1}$). High redshift events can also be identified by the presence of absorption below rest-frame 3000\AA\ due to heavy elements (Fe, Co, Ti) and highly ionised CNO-group elements \citep{Mazzali2016}. Unlike SNe Ia, the observed spectral features of a SLSN cannot necessarily be used to determine its exact phase. The spectral features observed are a function of the photospheric temperature of the SLSNe, and as SLSNe have been shown to evolve on a variety of different timescales \citep{Nicholl2017B}, this results in a much broader range of epochs over which a given spectral feature may be observed. As such, we do not require the epoch of the classification spectrum and that of the best matching spectral template to exactly coincide, allowing them to agree within $\sim\pm10$ days.

We assign the spectroscopic classification of SLSN-I for objects based upon spectroscopic behaviour similar to other literature SLSNe-I near peak, and where we can confidently rule out the presence of hydrogen (H$\beta$ or H$\alpha$ depending upon the redshift) at the time of observation. We do not place any luminosity restrictions in our spectroscopic classification. For one of the SNe, DES17E1fgl we identify the presence of H$\beta$ at an observed wavelength of 7388\AA, although the spectrum does not extend to redder wavelengths to confirm this classification with the addition of H$\alpha$ emission, so we thus classify this object to be \lq hydrogen-rich SLSN-like\rq\ (\lq SLSNe-II\rq).

Where the signal-to-noise ratio of the classification spectra is low ($\lesssim10$), or the match to spectral templates is visually poor, or the spectrum does not have adequate wavelength coverage to highlight these features which occur prominently in the rest-frame UV, we also consider the rankings of the best fitting spectral templates, as determined by \textsc{superfit}. We use the average rank of the templates of each spectroscopic class, whereby spectroscopic classes whose templates typically rank higher are used as a provisional classification. In such scenarios, we qualify the classification as \lq Silver\rq\ (as opposed to \lq Gold\rq\ where the match is more certain). This happened in only two cases. 

Where possible, we estimate spectroscopic redshifts using either host galaxy emission lines (H$\beta$, [\ion{O}{ii}] $\lambda$3727\,\AA, [\ion{O}{iii}] $\lambda\lambda$ 4959, 5007\,\AA) or narrow host absorption features such as \ion{Mg}{ii} $\lambda\lambda$ 2796, 2803\,\AA\ and/or \ion{Fe}{ii} $\lambda$2344\AA\ identified within the reduced spectra. In cases where the underlying host galaxy features could not be detected (6/22 events), we estimate the redshift from the best fitting spectral templates within \textsc{superfit}. 

\begin{table*}
	\centering
	\caption{SLSNe spectroscopically identified over the duration of the Dark Energy Survey. Redshifts are estimated from the host galaxy spectrum are marked \lq H\rq\ and those found using the features in SN spectrum are marked \lq S\rq. The precision of the measured redshifts varies depending upon the method through which they were derived (redshifts found from host spectra are typically better constrained). We also assign a classification \lq standard\rq\ of either Gold or Silver, depending upon our confidence in their classification.}
	\label{tab:DESsample}
	\begin{tabular}{l l l l l l l l} 
		\hline
		DES ID &  $\alpha$ & $\delta$ & $z$ & $z$ & Class &Standard& Rest-frame $\lambda$ \\
                &			&		 &  	& Source  &		&		&	Coverage \AA	 \\
		\hline
		DES13S2cmm	&02:42:32.83 	&-01:21:30.1	&	0.663 & H & SLSN-I &	Gold & 	2787	$\rightarrow$	5541		\\
		DES14C1fi	&03:33:49.80	&-27:03:31.6	&	1.302 & H & SLSN-I &	Silver & 	1931	$\rightarrow$	3840				\\
		DES14C1rhg	&03:38:07.27	&-27:42:45.7	&	0.481 & S & SLSN-I &	Gold & 	3130	$\rightarrow$	6222		\\
		DES14E2slp	&00:33:04.08	&-44:11:42.8	&	0.57 &  S & SLSN-I &	Silver & 	2952	$\rightarrow$	5870		\\
		DES14S2qri	&02:43:32.14	&-01:07:34.2	&	1.50 & S & SLSN-I &	Gold & 	1854	$\rightarrow$	3686		\\
		DES14X2byo	&02:23:46.93 	&-06:08:12.3	&	0.868 & H & SLSN-I &	Gold & 	2492	$\rightarrow$	4955		\\
		DES14X3taz	&02:28:04.46	&-04:05:12.7	&	0.608 & H & SLSN-I &	Gold & 	2883	$\rightarrow$	5731		\\
		DES15C3hav	& 03:31:52.17	&-28:15:09.5	&	0.392  & H & SLSN-I &	Gold & 	3335	$\rightarrow$	6630		\\
		DES15E2mlf	&00:41:33.40	&-43:27:17.2	&   1.861 & H & SLSN-I &	Gold & 	1621	$\rightarrow$	3222		\\
		DES15S1nog	&02:52:14.98	&-00:44:36.3	&	0.565 & H & SLSN-I &	Silver & 	2962	$\rightarrow$	5889			\\
		DES15S2nr	&02:40:44.62	&-00:53:26.4	&	0.220  & H & SLSN-I &	Gold & 	3799	$\rightarrow$	7554		\\
		DES15X1noe	&02:14:41.93	&-04:52:54.5	&	1.188 & H & SLSN-I &	Gold & 	2118	$\rightarrow$	4212			\\
		DES15X3hm	&02:26:54.96	&-05:03:38.0	&	0.860  & H & SLSN-I &	Gold & 	2492	$\rightarrow$	4955		\\
		DES16C2aix	&03:40:41.17	&-29:22:48.4	&	1.068 & H & SLSN-I &	Gold & 	2249	$\rightarrow$	4472			\\
		DES16C2nm	&03:40:14.83	&-29:05:53.5	&	1.998 & H & SLSN-I &	Gold & 	1546	$\rightarrow$	3074			\\
		DES16C3cv	&03:27:16.71	&-28:42:45.9	&	0.727 & H & SLSN-I &	Silver & 	2684	$\rightarrow$	5336		 \\
		DES16C3dmp	&03:31:28.35	&-28:32:28.3	&	0.562 & H & SLSN-I &	Gold & 	2961	$\rightarrow$	5888			\\
		DES16C3ggu	&03:31:12.00	&-28:34:38.7	&	0.949 & H & SLSN-I &	Gold & 	2377	$\rightarrow$	4726			\\
        DES17E1fgl	&00:32:09.62	&-42:38:49.3	&	0.52  & H & SLSN-II & 	Gold &	2967	$\rightarrow$	5899			 \\
        DES17X1amf	&02:17:46.70 	&-05:36:01.0	&	0.92  & S & SLSN-I & 	Gold &	2414	$\rightarrow$	4799			\\
        DES17C3gyp	&03:27:51.87 	&-28:23:44.3	&	0.47  & S & SLSN-I &	Silver &  3152	$\rightarrow$	6268		\\
        DES17X1blv	&02:20:59.64 	&-04:29:00.8	&	0.69  & S &	SLSN-I &	Gold &	2742	$\rightarrow$	5452		\\
		\hline
	\end{tabular}
\end{table*}

\begin{figure*}
	\centering
	\includegraphics[scale=0.40]{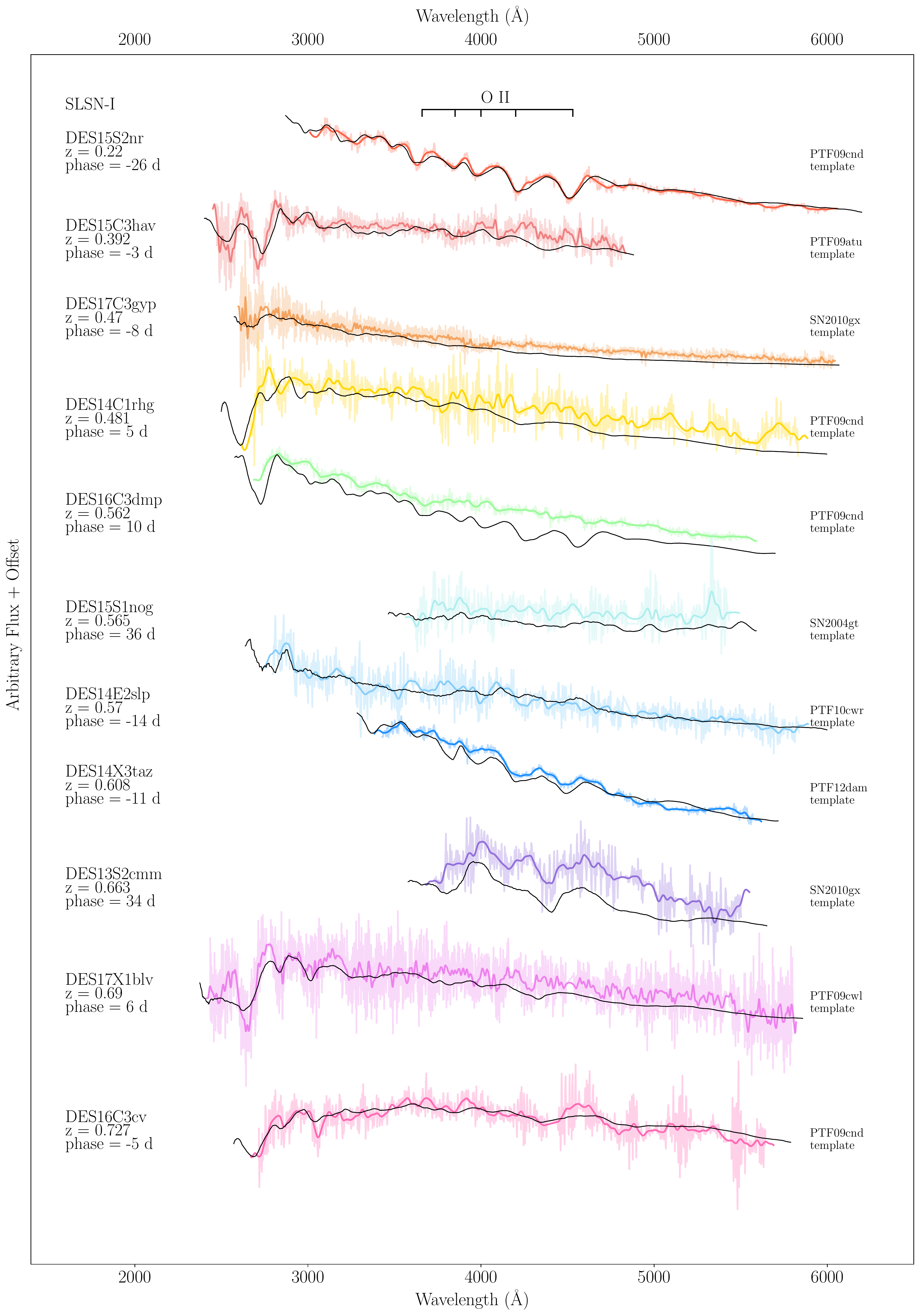}
    \caption{Spectra used for the spectroscopic confirmation of DES SLSN-I candidates, with observational details given in Table~\ref{tab:spec_obs}. The DES spectra are shown unbinned and binned by 5\,\AA\ bins to increase the clarity of the features. The phase provided corresponds to the phase at which the classification spectrum was taken in the SN rest frame, not that of the spectral template. The black lines show the smoothed best fitting spectral template used to classify the spectrum for each SN (see \citealt{Quimby2018} for more template details). We determine spectroscopic redshifts from either the presence of host galaxy absorption lines, or from fitting the observed SN features to those of established SLSNe within the literature. Table \ref{tab:DESsample} contains details of the reshift source and classification quality for each event. Here we highlight the \ion{O}{ii} absorption lines seen in the optical spectra around 4000\AA. }
    \label{fig:DES_Spec_i}
\end{figure*}
\renewcommand{\thefigure}{\arabic{figure}{ Cont}}
\addtocounter{figure}{-1}

\begin{figure*}
	\centering
	\includegraphics[scale=0.40]{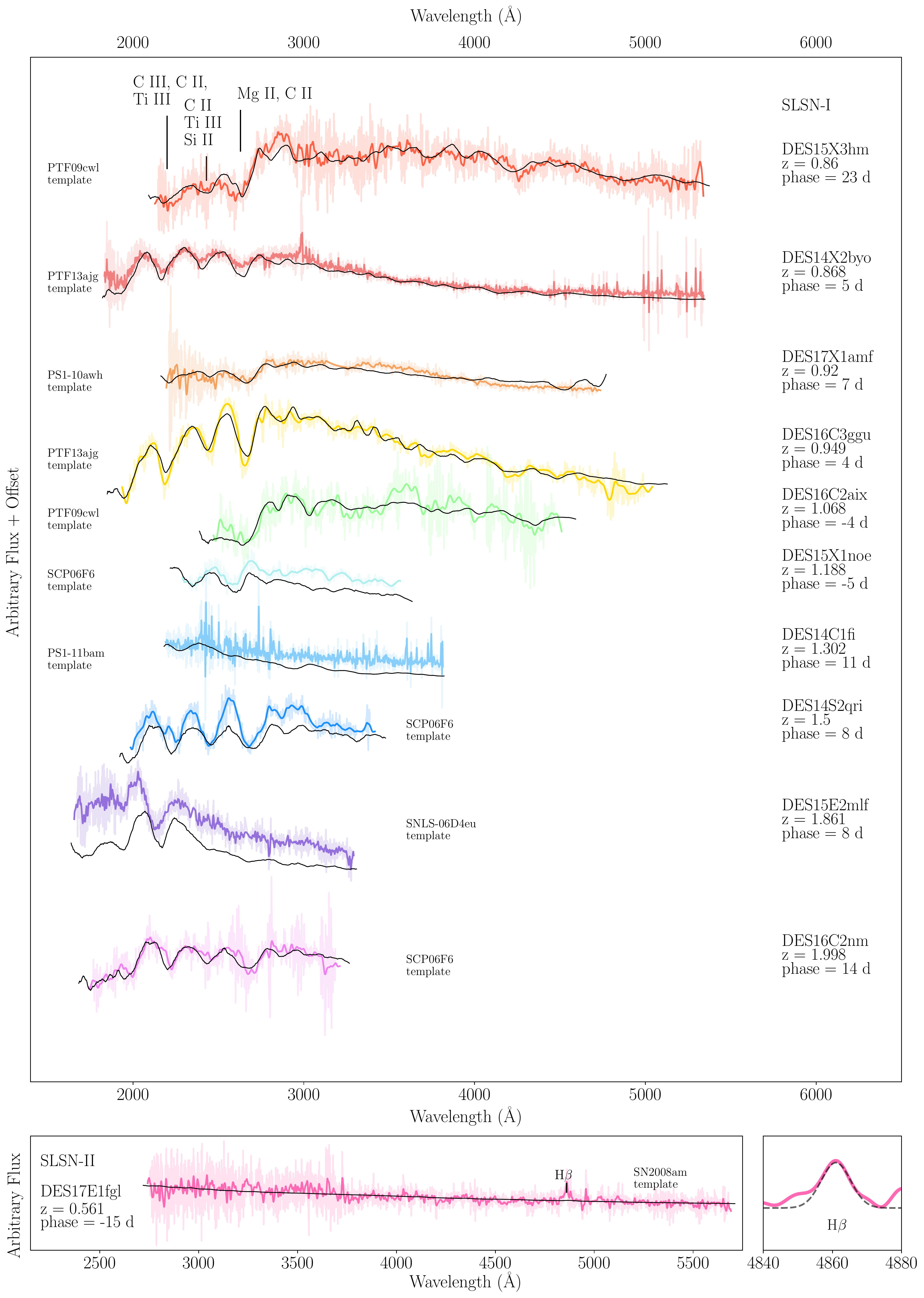}
    \caption{Figure 1 continued. We highlight some of the typical spectroscopic features observed within SLSN-I at higher redshift -- namely blended absorption from C, Mg, Ti and Si lines below 3000\AA. We also show the spectra used to confirm our SLSN-II candidate (lower panel), with the H$\beta$ line used to classify this event highlighted in the lower right panel. Table \ref{tab:DESsample} contains details of the reshift source and classification quality for each event.}
    \label{fig:DES_Spec}
\end{figure*}

\renewcommand{\thefigure}{\arabic{figure}}

\begin{figure}
	\includegraphics[width=\columnwidth]{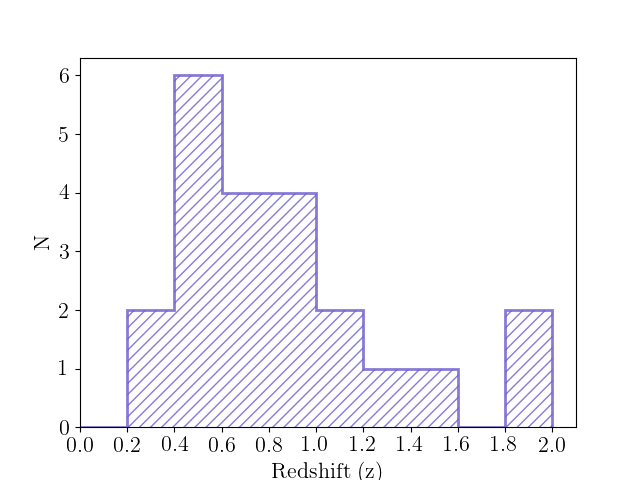}
    \caption{The redshift distribution of the DES SLSNe, grouped into bins of $z=0.2$. The median redshift of the sample is $z=0.86$.}
    \label{fig:DES_redshifts}
\end{figure}

From these data, we have spectroscopically identified a sample of 22 SLSNe\footnote{The remaining 10 objects triggered as candidate SLSNe did not contain sufficient signal in their spectra to procure a secure classification.}. We present our classification spectra in Fig.~\ref{fig:DES_Spec_i}, alongside the best fitting spectral template for each object, and details of all spectroscopic observations can be found in in Table~\ref{tab:spec_obs}. Basic information on each event (including the classification quality) can be found in Table~\ref{tab:DESsample}. Our final sample spans a broad redshift range of $0.220\leq~z~\leq1.998$ (Fig.~\ref{fig:DES_redshifts}). The deep imaging capability of DES enables the detection of both local \lq fainter\rq\ SLSNe, as well as their higher redshift counterparts, including some of the most distant spectroscopically confirmed SNe to date \citep{Pan2017,Smith2018}. This broad redshift range consequently results in a wide range of rest-frame wavelengths being probed from object to object. We list the rest-frame wavelengths of the DECam filters for each object in Table~\ref{tab:DESsample}. 

\subsubsection{Light Curves}
\label{sec:lightcurves}

Photometric measurements of all DES SLSNe were made using the pipeline discussed by \citet{Smith2016}, which has also been extensively used in the literature \citep[e.g.,][and references therein]{Firth15}. This pipeline performs classical difference imaging by subtracting a deep template image (typically 4-6 times deeper than a single photometric epoch alone) from each individual SN image to remove the host-galaxy light using a point-spread-function (PSF) matching routine. SN photometry is then measured on the difference images using a PSF fitting technique. The $g,r,i$ and $z$ light curves for every confirmed DES SLSNe are presented within Fig.~\ref{fig:DES_LC}, and the photometry is given in Appendix~\ref{appendix_photom}. All reported magnitudes are corrected for Milky Way extinction following \cite{Schlafly2011}.   

\begin{figure*}
	\centering
	\includegraphics[scale=0.36]{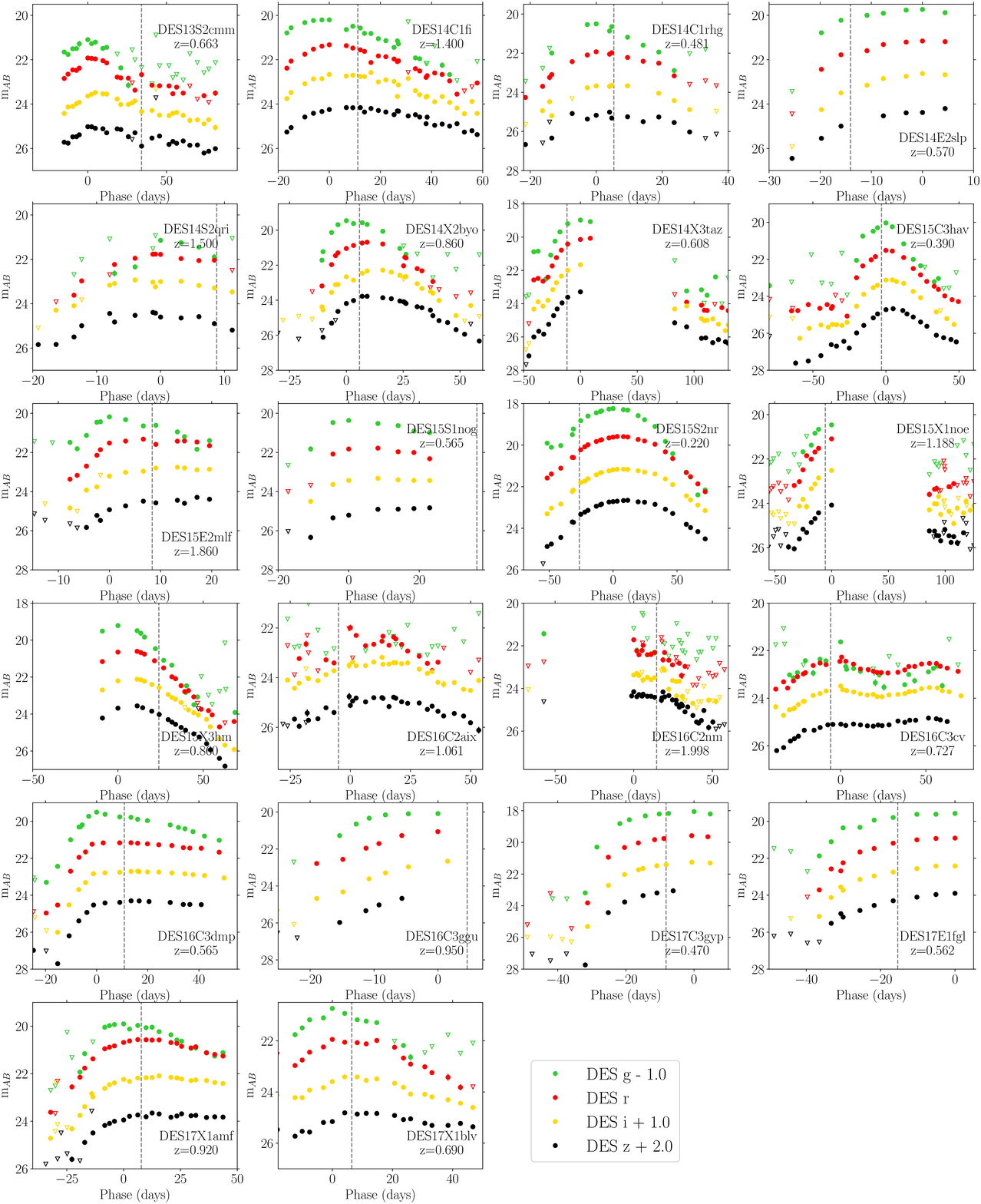}
    \caption{The DES $g,r,i,z$ light curves of our sample of SLSNe. All magnitudes are in the AB system and are corrected for Milky Way extinction. All light curves show the phase of the SNe with respect to the peak in the observer-frame $g$-band (or $r$-band for objects with no $g$-band flux). Unfilled triangles are non-detections reported as 3-$\sigma$ upper limits. For clarity the $g,i$ and $z$ bands are offset by $-1$, 1 and 2 magnitudes respectively. Dashed grey lines indicate the phases at which classification spectra were obtained.}
    \label{fig:DES_LC}
\end{figure*}

\section{Light curve interpolation}
\label{sect:mod}

Our next task is to develop the framework to calculate SN brightnesses at any epoch by interpolating the DES observations. This is required to estimate the peak brightnesses of the SNe, as well as to estimate bolometric luminosities. In this paper we use Gaussian Process techniques to perform the light curve interpolation. 

Gaussian Processes (GP) are a generalised class of functions, which may be used to model the correlated noise within time-series data. They represent a distribution over the infinite number of possible outputs of a function, such that the distribution over any finite number of them is a multivariate Gaussian. GP functions consist of a mean function ($\mu(x)$) and a \lq kernel\rq\ ($k(x,x^{\prime})$), which is a simple function used to describe the covariance between data points $x$ and $x^{\prime}$.
\begin{equation}
f(x) = GP\big(\mu(x),k(x,x^{\prime})\big)
\end{equation}

GPs may be considered a generalisation of a Gaussian distribution; at each point $x$, the reconstructed function $f(x)$ may be described by a normal distribution. Neighbouring points are not independent, but are related by some predetermined covariance function, given by the kernel. The use of GP techniques allows the user to marginalise over systematic sources of noise within a data, which might otherwise not be captured in an astrophysical model.

At each point in the reconstructed function, the precise value of the resultant value of the function is unknown, but does lie within a normal distribution \citep{Rasmussen2006}. The mean function $\mu(x)$ describes the mean value drawn from this distribution of possible answers, while the covariance function used within the GP determines the form of the relationship between surrounding data points. The covariance function used here depends upon two hyper-parameters; the variance of the signal within the function, and the characteristic scale length over which any significant signal variations occur. For time-series data, these two hyper-parameters respectively become the uncertainties in measured flux and the timescale over which significant changes occur within the data. The functional form of the covariance function or \lq kernel\rq\ used may be selected/constructed such that it reflects any periodic tendencies within the data. 

The final likelihood function of a GP is a multivariate Gaussian with $N$ dimensions, but in which the measurements are dependent as dictated by a covariance matrix, which absorbs any systematics that are unknown to the user. In order to produce the best interpolation between data points, the hyper-parameters of the kernel may be optimally fit for. 

Here we interpolate the multi-colour light curves of the DES SLSNe using GP fitting. To do this we utilise the {\sc{python}} package {\sc{george}} \citep{Ambikasaran2014} and optimally fit the hyper-parameters of a Matern 3/2 kernel independently for each of the $g,r,i,z$ light curves. A Matern 3/2 kernel is mathematically similar to the more familiar squared exponential function (but with a narrower peak) and has the form
\begin{equation}
k(r) = \sigma^{2}\bigg(1+\frac{{\sqrt{3}r}}{l}\bigg)\exp{\Bigg(\frac{-{\sqrt{3}r}}{l}\Bigg),}
\end{equation}
\noindent
where $\sigma$ is the uncertainty of an observation, $r$ is the separation between observations and $l$ is the characteristic scale length over which variations in the data occur.

We chose this kernel form as we find the sharper peak results in greater flexibility within the final function over short timescales, which acts to best capture the visual form of the SLSN light curves. Prior to interpolation, we perform a gradient based optimization to determine the best fitting hyper-parameters for the kernel in each photometric band. We interpolate the observations using GPs over regions where we have SN detections in multiple bands. For SN detected over multiple seasons, the interpolation is carried out over the entire duration of the transient, although we note that between observing epochs the interpolation is highly unconstrained. 

The independent fitting of the $g,r,i,z$ light curves fails to take into account any wavelength overlap between the DECam photometric filters. Given the fairly even sampling of the survey across all 
4 filters, it is possible to improve constraints upon the light curve behaviour through the use of a 2-dimensional kernel in which the wavelength scale is also accounted for. However, a 2-dimensional kernel is beyond the aims of this study.

\begin{figure*}
	\centering
	\includegraphics[scale=0.36]{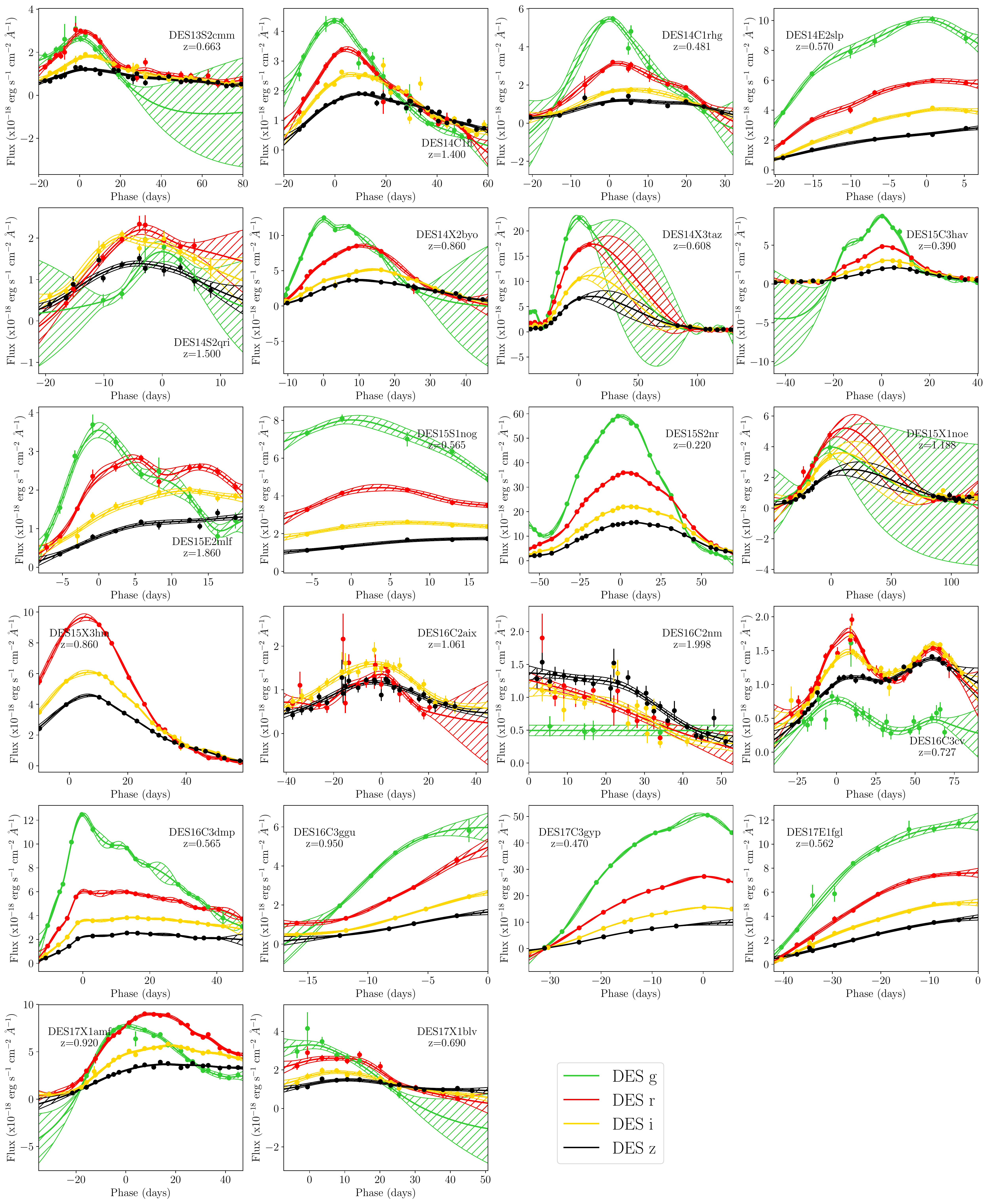}
    \caption{The $g,r,i$ and $z$ Gaussian Process interpolated light curves of DES SLSNe using an optimised Matern 3/2 kernel. Limits are not shown here. All light curves show the phase of the SNe with respect to the peak in the observer-frame $g$-band (or $r$-band for objects with no $g$-band flux). These interpolated light curves are used in Sections~\ref{sect:mod}, and \ref{sect:lc_res} to determine the light curve parameters and to fit physical models to the data.}
    \label{fig:GP_LC}
\end{figure*}

For each GP fit, we present the mean and 1$\sigma$ uncertainties in Fig.~\ref{fig:GP_LC}, with the DES photometry. With each optimised kernel, the interpolated light curves have less flexibility in areas with a greater data density (where the data points are well constrained), as the underlying trend can be specified at a higher confidence. Conversely, interpolations during large gaps in the data are more uncertain.

Visual inspection of the final interpolations in Fig.~\ref{fig:GP_LC} shows that they capture the evolution of the SLSN well. We do observe some \lq ballooning\rq\ of the uncertainties between well constrained data points where the gap between observations is large (for example, the light curve of DES16C3dmp shows such a feature between observations at $+18$ and $+28$ days). This is a natural consequence of the GP interpolation, as the optimised kernel determines the coherence between observations. High cadence, well-constrained observations result in fewer degrees of freedom for the fit between observations. More sparsely spaced, but well constrained observations in the light curve will generate more uncertainty in the final fit. 

The interpolated fits provided by the GP provide photometry estimates at any epoch, regardless of the spacing in the observations. This allows an estimation of the peak epoch of the SN, as well as its evolutionary timescales. We define these light curve properties for each in SN in its bluest observed band (either DES $g$ or $r$, depending on the redshift), using the GP interpolated fits to determine the peak MJD, rise and decline times. These properties are presented in Table~\ref{tab:slsnphot}.  

\section{Modelling the Light curves}
\label{sect:lc_res}

In order to investigate the emerging diversity of physical characteristics attributed to SLSNe within the literature, it is necessary to apply models to the observed light curves such that their the rest frame properties can be determined. We fit the following physical models to the interpolated DES SLSN light curves; a modified black body and a magnetar model. 

\subsection{A modified black body model}

For simplicity in determining luminosities and k-corrections for our light curves, we use an approximate spectral energy distribution (SED) to generate synthetic photometry and fit this to the data. While SLSNe-I present largely featureless spectra in relation to other SN classes, fairly well represented by a black body at optical to NIR wavelengths \citep[e.g.][]{Nicholl2017B}, they deviate from a black body SED at ultraviolet (UV) wavelengths, with significant absorption below $\sim$\textless 3000\,\AA. These features, largely attributed to absorption by heavy elements \citep[see ][for example line identification]{Mazzali2016}, will dominate the observable optical light curves of the higher redshift SLSNe found within DES. We therefore adopt a modified black body model for our SLSN SEDs.

\begin{figure*}
	\centering
	\includegraphics[scale=0.52]{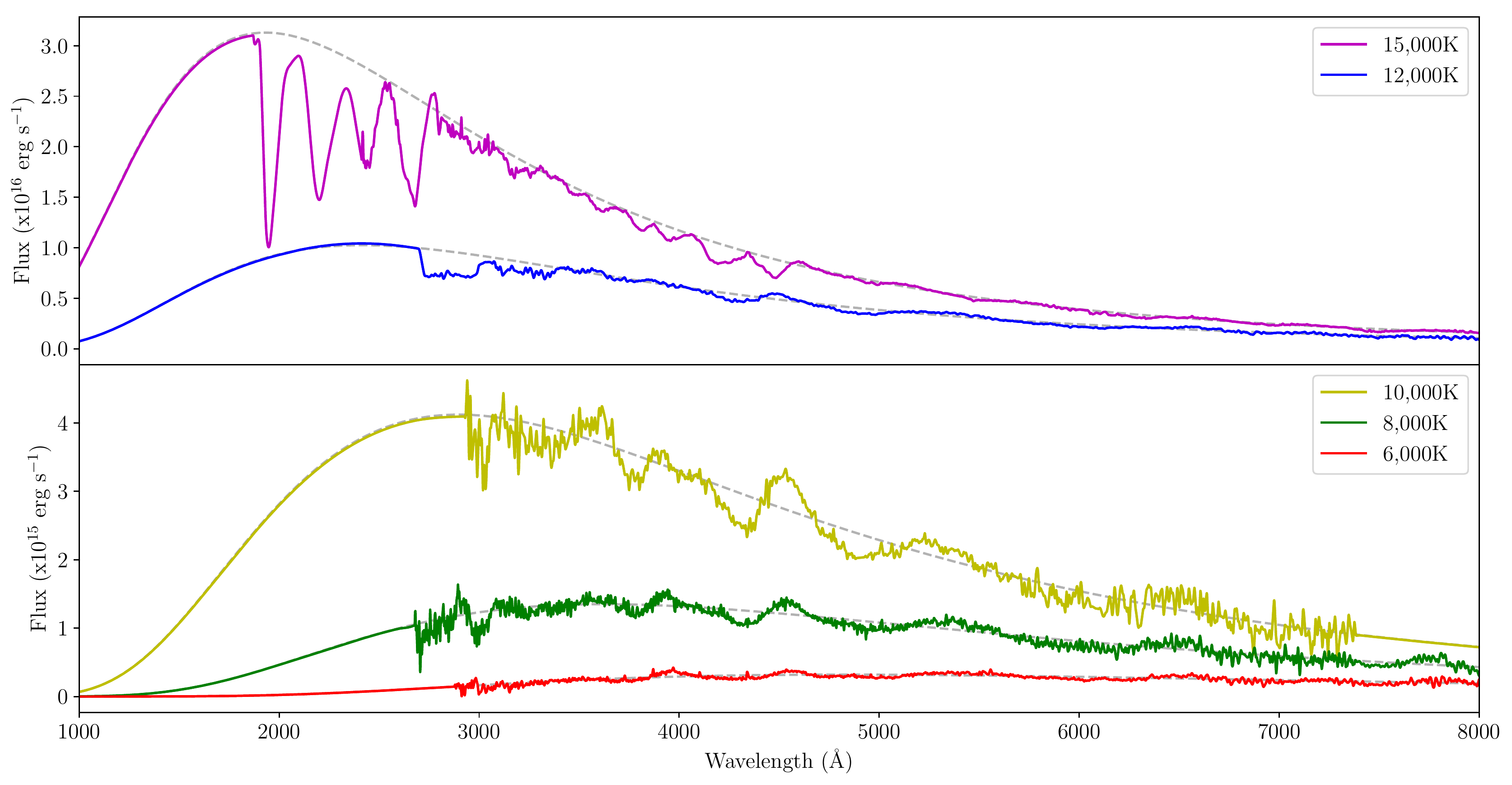}
    \caption{The spectroscopic templates of SLSNe at temperatures of {\textit{(bottom panel)}}~ 6000, 8000, 10000, and {\textit{(top panel)}}~ 12000 and 15000\,K, (corresponding phase$>35$ days, $22<$phase$<35$ days, $15<$phase$<22$ days, $10<$phase$<15$ days and $<10$ days respectively) derived from the averaged spectra of literature SLSNe with rest-frame UV coverage \citep[as per][]{Inserra2017Euclid}. Grey dashed lines indicate the SED of an unabsorpbed black body at the same temperature. We fit the featureless continuum regions of these spectra in 50\AA\ wide bins with Planck's law, and thus determine the ratio between the template and the black body SED. These final templates are applied to the DES SLSN black body SEDs to approximate the strength of UV absorption as a function of wavelength and temperature.}
    \label{fig:UV_absorption}
\end{figure*}

We construct our modified SEDs in the following manner. We use empirical templates of UV absorption based upon 110 rest-frame spectra of SLSNe-I \cite[as per][]{Inserra2017b}, from which we estimate the variation of UV absorption at temperatures of 15000\,K, 12000\,K, 10000\,K, 8000\,K and 6000\,K (corresponding to the following approximate phases respectively; $<10$ days, $10<$phase$<15$ days, $15<$phase$<22$ days, $22<$phase$<35$ days and $>35$ days from peak) \footnote{At temperatures below 6000\,K, the spectrum is assumed to have little UV absorption and is therefore treated as a normal black body}. These spectroscopic templates are presented within Fig.~\ref{fig:UV_absorption}. Due to a lack of spectroscopic information bluewards of $\sim2000 - 2500$\,\AA\ depending upon the temperature/phase of the SN, we model our SEDs shortwards of these wavelengths as that of a black body. This assumption may lead to an overestimation of flux in bolometric calculations (if a blueward bolometric correction were included), particularly around maxiumum light (i.e. the 15,000K template), when the SED peaks around this region. However, it should not impact our k-correction for blue photometric bands with wavelengths
greater than $\sim$3000\AA, and as only 2 objects within our sample are probed by the DES photometry at wavelengths bluer than the template boundaries (DES16C2nm and DES15E2mlf), this should not significantly impact the results presented here. 

We then follow the methodology of \cite{Prajs2016}, fitting the featureless regions of the continuum within the templates in 50\,\AA\ wide bins with Planck's law. From this we determine the ratio between the template and the black body continuum that we use as measure of the strength of the absorption features as a function of wavelength. The strength of the UV absorption is highest around the peak of the SN (i.e., when the photosphere is hottest), and generally decreases in strength as the SN cools.  At longer wavelengths, we assume no additional absorption and therefore adopt a standard Planck black body for the SED. 

Finally, for any k-corrections to other photometric filters we use the \lq mangling\rq~
technique of \cite{Hsiao2007} to colour correct our synthetic SEDs. We do this by fitting a spline function to our photometry and then applying this function to the SED, thus ensuring that the SED replicates the colours of our $g,r,i,z$ photometry.

\begin{figure*}
	\centering
	\includegraphics[scale=0.6]{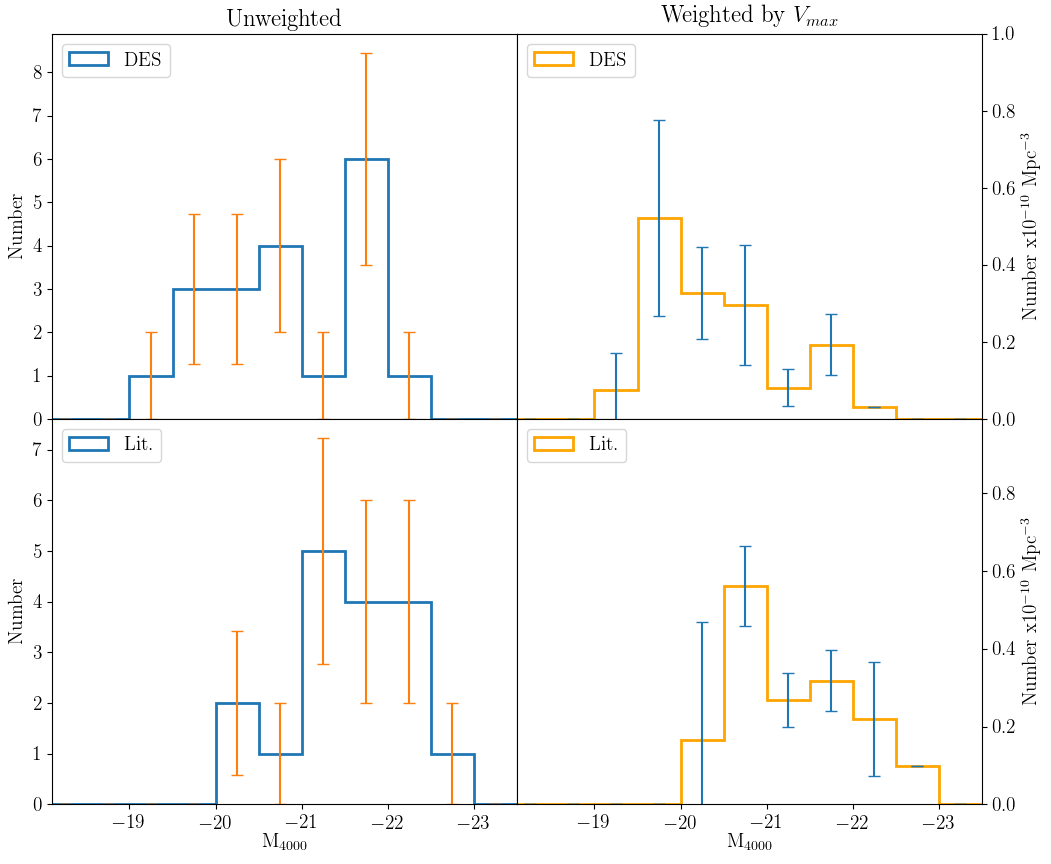}
    \caption{SLSN-I luminosity functions in the synthetic 4000\,\AA\ band \citep{Inserra2014}. The panels on the left show the unweighted distributions of peak luminosities for DES SLSNe-I (top panels), and for spectroscopically-confirmed SLSNe-I from the literature (lower panels). The panels on the right show the same distributions but weighted by $1/V_\mathrm{max}$, the maximum volume out to which each event could have been classified. Here the sensitivity of DES highlights the presence of an increased population of events peaking at fainter luminosities.}
    \label{fig:DES_lum_func}
\end{figure*}

\subsubsection{The SLSN-I luminosity function}
\label{Sec:luminosityfunction}

We fit our modified black body model to the GP interpolated $g,r,i,z$ light curves, with the peak temperatures and radii from these fits within Table~\ref{tab:slsnproperties}. We then use these SEDs to  determine the absolute peak luminosities of the SLSN light curves within the artificial 4000\,\AA\ band first defined in \citet{Inserra2014}. This band was selected for the purposes of standardization as it encompasses a region of a typical SLSN spectrum which is usually dominated by featureless continuum for up to $\sim30$ days after peak. We thus determine these magnitudes using the best fitting modified black body spectrum to the observed photometry of at the SN peak within the SN rest-frame. These peak luminosities are presented within Table~\ref{tab:slsnphot}, and the distribution of peak magnitudes in Fig.~\ref{fig:DES_lum_func}.

\begin{table*}
	\caption{The derived photometric properties of the DES SLSN sample. Rise times ($\tau_{\mathrm{rise}}$) are measured using the GP interpolations, from the epoch of the last non-detection to peak brightness, and vice-versa for decline times ($\tau_{\mathrm{decline}}$).}
	\centering
	\begin{tabular}{l l l l l l l} 
		\hline
		DES ID 	& MJD$_{peak}$ & Peak L$_{\mathrm{Bol}}$ & $\tau_{\mathrm{rise}}$ & $\tau_{\mathrm{fall}}$ & Peak M$_{\mathrm {4000}}$ &  Peak M$_{\mathrm {5200}}$ \\
  & (J2000) 		&erg~s$^{-1}$	& (days)    		& (days) 			&    \\
		\hline
		DES13S2cmm	& 	 56562.4	&	2.043e+43	& 	20.0	&	64.1 &   -19.63 $\pm$	0.46	&	-19.96 $\pm$	0.42 \\
		DES14C1fi	& 	 56920.8	&	1.481e+44	& 	22.0	&	59.9 &   -21.96 $\pm$	0.30	&	-22.36 $\pm$	0.27 \\
		DES14C1rhg	& 	 57010.9	&	9.015e+42	& 	20.9	&	30.0 &   -19.58 $\pm$	1.01	&	-19.71 $\pm$	0.68 \\
		DES14E2slp	& 	 57040.6	&	3.067e+43	& 	19.9	&	8.02 &   -20.51 $\pm$	0.07	&	-20.48 $\pm$	0.06 \\
		DES14S2qri	& 	 57050.4	&	9.196e+43	& 	7.6		&	22.4 &   -21.57 $\pm$	0.19	&	-21.59 $\pm$	0.18 \\
		DES14X2byo	& 	 56944.7	&	8.991e+43	& 	10.0	&	32.2 &   -21.67 $\pm$	0.13	&	-21.64 $\pm$	0.12 \\
		DES14X3taz	& 	 57077.5	&	1.069e+44	& 	29.1	&	10.9 &   -21.72 $\pm$	0.20	&	-21.80 $\pm$	0.18 \\
		DES15C3hav	& 	 57339.1	&	1.057e+43	& 	19.9	&	40.0 &   -19.57 $\pm$	0.60	&	-19.69 $\pm$	0.47 \\
		DES15E2mlf	& 	 57348.8	&	2.137e+44	& 	8.0		&	19.9 &   -21.99 $\pm$	0.43	&	-21.79 $\pm$	0.40 \\
		DES15S2nr	& 	 57318.9	&	1.760e+43	& 	48.9	&	68.2 &   -20.28 $\pm$	0.16	&	-20.36 $\pm$	0.14 \\
		DES15X1noe	& 	 57423.7	&	1.106e+44	& 	48.8	&	125.0 &  -23.37 $\pm$	3.17	&	-24.16 $\pm$	2.74 \\
		DES15X3hm	& 	 57229.4	&	1.410e+44	& 	- 		&	66.9 &   -22.00 $\pm$	0.15	&	-21.86 $\pm$	0.14 \\
		DES16C2aix	& 	 57715.1	&	1.257e+43	& 	26.0	&	43.1 &   -20.81 $\pm$	0.60	&	-20.99 $\pm$	0.54 \\
		DES16C2nm	& 	 57620.0	&	1.062e+44	& 	- 		&	52.0 &     -22.82 $\pm$ 0.67			& -23.18 $\pm$ 0.61	\\
		DES15S1nog	& 	 57365.2	&	1.947e+43	& 	11.7	&	16.3 &   -20.32 $\pm$	0.50	&	-20.27 $\pm$	0.46 \\
		DES16C3cv	& 	 57665.5	&	1.348e+43	& 	42.0	&	64.9 &   -19.18 $\pm$	0.40	&	-19.86 $\pm$	0.35 \\
		DES16C3dmp	& 	 57725.2	&	2.769e+43	& 	10.9	&	42.0 &   -20.66 $\pm$	0.34	&	-20.54 $\pm$	0.32 \\
		DES16C3ggu	& 	 57794.2	&	6.531e+43	& 	15.0	&	-	  &  -21.09 $\pm$	0.37	&	-20.92 $\pm$	0.56 \\
        DES17E1fgl	&	 58136.1	&	3.195e+43	& 	42.0	&	-	  &  -20.58 $\pm$	0.18	&	-20.66 $\pm$	0.16 \\
        DES17X1amf	&	 58049.6	&	1.037e+44	& 	24.2	&	41.9 &   -21.61 $\pm$	0.42	&	-21.79 $\pm$	0.38 \\
		DES17C3gyp	& 	 58049.6	&	7.787e+43	& 	24.2	&	41.9 &   -21.69 $\pm$	0.14	&	-21.63 $\pm$	0.13 \\
        DES17X1blv	& 	 58049.6	&	1.724e+43	& 	24.2	&	41.9 &   -20.04 $\pm$	0.02	&	-20.16	$\pm$	0.02 \\
        \hline
		\label{tab:slsnphot}
	\end{tabular}
 \end{table*}

For comparison we also determine the 4000\,\AA\ luminosity function of other spectroscopically classified SLSNe-I from the literature. To do this we take the previously published photometry and perform GP interpolation in the same way as for the DES objects. We select the literature sample based upon the following criteria:
\begin{itemize}
\item Each SN must have been spectroscopically classified as a SLSN-I or \lq SLSN-I like\rq,
\item Each SN must have been observed in a minimum of three different photometric bands for better parameterization when fitting the SEDs,
\item Each observed band must have a minimum of five epochs of data.
\end{itemize}
We process the interpolated literature light curves in an analogous manner to the DES SLSNe, using the modified black body function described previously. We perform a volumetric correction to both our samples of SLSNe-I using the maximum volume over which the SNe could have been detected ($V_{\mathrm{max}}$). We use a limiting magnitude for spectroscopic confirmation of $m_{R}\sim23.5$\,mag\footnote{This limit is representative of the depth achievable within a standard ToO 1 hour observation with VLT X-shooter under good observing conditions.} to determine the maximum volume out to which the particular SLSN could have been spectroscopically identified. In Fig.~\ref{fig:DES_lum_func}, we observe a clear difference in the distributions of transient luminosities between the two samples, with the depth of DES more frequently identifying fainter SLSNe, with four SLSNe events peaking at $M>-20$, while the observational bias for spectroscopic follow up of brighter objects within literature SLSN samples skews this distribution to higher luminosities. This trend persists in the volume-corrected distributions, where we see the majority of DES objects lie fainter than the arbitrary $M<-21$ limit originally used to classify SLSNe.

\begin{figure*}
	\includegraphics[scale=0.6]{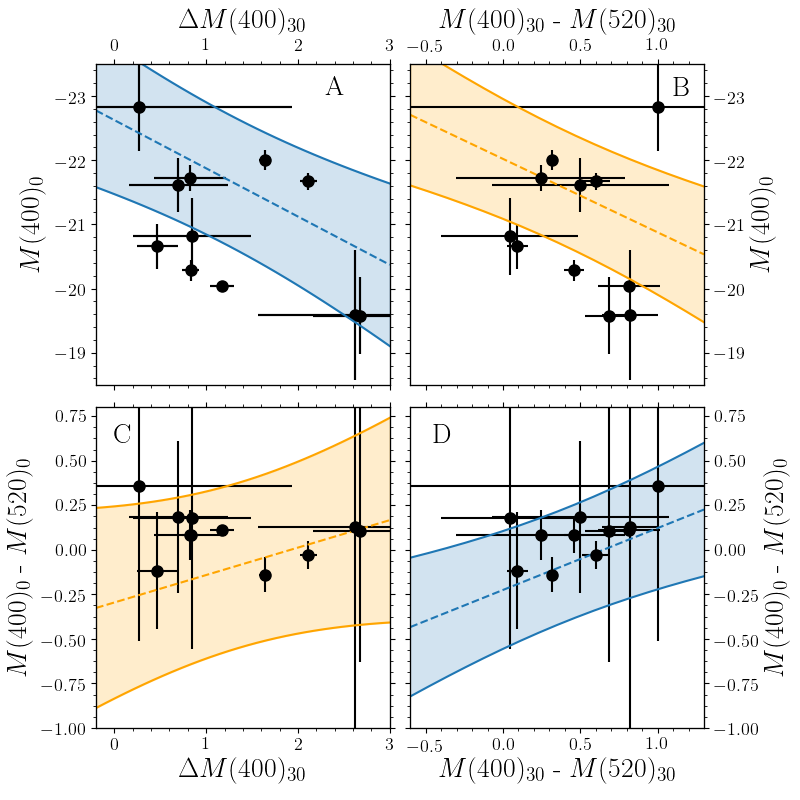}
    \caption{The \lq 4OPS\rq\ SLSN parameter space introduced by \citet{Inserra2018}, showing the statistical relations in the luminosity and colour evolution of SLSNe. Panel A (top left) shows the relationship between peak luminosity and luminosity at $+30$\,days in the 4000\,\AA\ band; Panel B (top right) the relationship between the $4000-5200$\,\AA\ colour and peak 4000\,\AA\ luminosity; Panel C (bottom left) the relationship between the luminosity at 30 days and the colour at peak and finally; Panel D (bottom right), the relationship between colour at peak and colour at 30 days. The orange and blue shaded regions highlight the 3$\sigma$ confidence bands within which this parameter space has been defined for previous samples of SLSNe. Only 11 of the 22 DES SLSNe possess information at $+30$ days from maximum light, such that they can be included in the 4OPS plot. Due to the lack of data and extensive extrapolation around peak for DES15X1noe, we do not include this object within the 4OPS plot.}
    \label{fig:DES_4OPS}
\end{figure*}

This broad distribution in peak luminosities exhibited within the DES SLSN-I sample brings the congruence amongst other SLSNe into question. We therefore compare their luminosity and colour evolution using the \lq 4OPS\rq\ parameter space defined in \cite{Inserra2018}, which compares the colour and evolution in the artificial 4000\,\AA\ and 5200\,\AA\ bands which have shown to be significantly correlated for other SLSNe in the literature.  We determine the 5200\,\AA\ light curves of the DES SLSNe in the same manner as their 4000\,\AA\ light curves, and using the luminosities in these bands at both peak and $+30$ day phases (where possible), place them in the 4OPS parameter space in Fig.~\ref{fig:DES_4OPS}.  

The 4OPS parameter space gives some insight into the physical properties of the SN ejecta, for instance a correlation between colour at two different phases suggests a link in temperature or radius of the ejecta during these epochs \citep{Inserra2018}. Within the paradigm of the magnetar model of SLSN production, these relations have been interpreted as the energy from the magnetar being injected at the same epoch for all SLSNe \citep{Inserra2018}, as this would naturally create a similar timescale for the rise of the SN and the diffusion timescale through the ejecta. 

Whilst the DES SLSNe fall within the colour-evolutionary space defined by other SLSNe, they push the boundaries of some of these statistical relationships, with a significant fraction of the sample falling on average $\sim1$ magnitude fainter at peak than the 3$\sigma$ space statistically occupied by other SLSN events of the same decline rate. Many events also evolve on more rapid timescales than the 3$\sigma$ parameter space originally defined with literature events too. This is likely a reflection of the softer selection criteria designed in selection of the DES SLSN sample. We do not find any significant correlation between the peak luminosities and peak colours of these events, which does not suggest a strong link between the peak luminosity and photospheric temperature for all events within the sample.

This broad spread of luminosities could be due to a range of injection times of the magnetar energy with respect to the SN. A delayed injection would result in typically fainter peak magnitudes, due to a combination of the lagging magnetar energy diffusing through the SN ejecta behind the main SN shockwave, and a reduced energy input from the magnetar due to a loss in its rotational energy in the period between core collapse and energy injection \citep{Woosley2018}. We explore the magnetar model in more detail within Section \ref{sect:magnetar}. 

\subsubsection{Bolometric Lightcurves}

\begin{figure*}
	\includegraphics[scale=0.6]{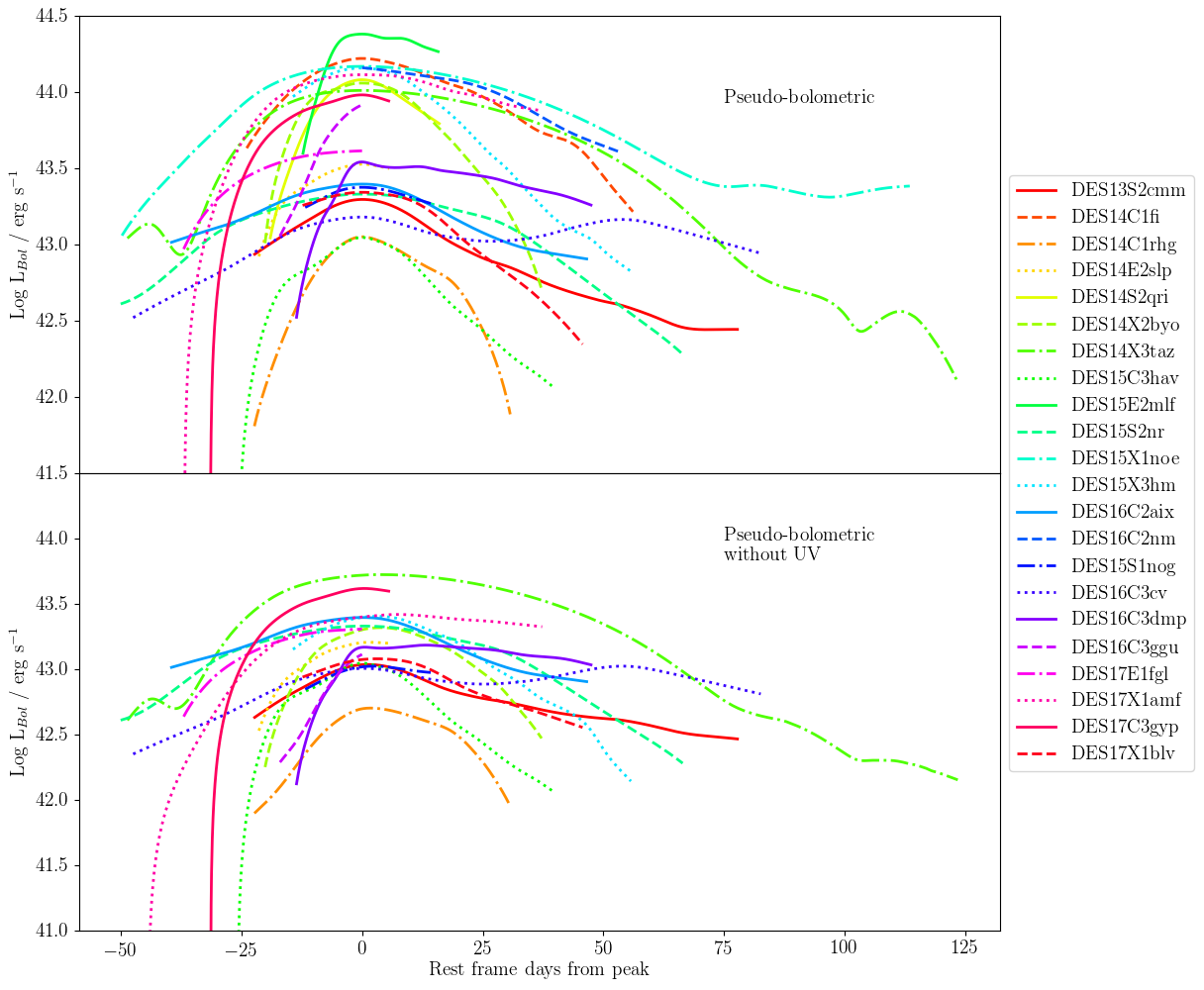}
    \caption{{\textit{Top:}} The pseudo-bolometric light curves of DES SLSNe constructed through trapezoidal integration of the $g,r,i,z$ photometry and applying a bolometric correction redwards of the most red observed band to 25000\AA. The large range in apparent peak energies, spanning almost 2\,dex, could reflect a large spread in initial explosion energies, or imply an efficiency factor in the transferring the initial explosion energy to the SN. {\textit{Bottom:}} Pseudo-bolometric light curves with a consistent blue cut-off above $\sim$3800\AA. We still see a considerable dispersion in radiated energies when considering similar wavelength ranges in the pseudo-bolometric light curves.}
    \label{fig:DES_bol_lc}
\end{figure*}

We determine the pseudo-bolometric lightcurves of the SLSNe via trapezoidal integration of the flux in each band at every epoch to obtain a lower limit on the total emitted flux, applying a small red bolometric correction from integration of the best fitting modified black body curve redwards of the most red observed band to 25,000\AA. Whilst this correction is typically small ($\sim0.1$ per cent around peak for most objects), this contribution can become more significant ($\sim1$ per cent) at late times when the photosphere is cooler. This approach does loose information in the UV, whose contribution to the SED is more significant at early times. However, the rest frame wavelength range probed by the DES photometry does probe the near-UV for the vast majority of objects in this SLSN sample, allowing us to better encapsulate the bolometric behaviour of these events at early times. We present the pseudo-bolometric light curves in the upper panel of Fig.~\ref{fig:DES_bol_lc}.

The large observed spread in peak energies of the DES SLSNe is reinforced within Fig \ref{fig:DES_bol_lc}, where we observe an apparent $\sim2$\,dex spread in pseudo-bolometric peak flux. This range of peak pseudo-bolometric fluxes approaches much lower energies than has been observed within other SLSN samples \citep{DeCia2017,Lunnan2017}, with several events peaking at $\lesssim10^{43}$\,erg\,s$^{-1}$, comparable to the peak energies achieved by SNe Ia. This spread in peak luminosities between $\sim10^{43}$\,erg\,s$^{-1}$ and $\sim10^{44}$\,erg\,s$^{-1}$ could be a result of a very flexible progenitor set up (for instance, energy injection from magnetars with a range intrinsic properties), or a reflection of multiple energy sources producing these spectroscopically similar transients. 

Given the high-redshift of many objects within our sample, it is possible that the observed spread we see in integrated luminosities is a result of the higher UV contributions from these more distant SLSNe, given the redshift-dependent wavelength range being probed. We therefore present pseudo-bolometric lightcurves with a consistent blue cut-off above $\sim$3800\AA (the bluest wavelength probed by the DES observations for the lowest redshift SN in our sample) in the lower panel of Fig.~\ref{fig:DES_bol_lc}, such that we are comparing a similar wavelength range over the entire sample. To avoid excessive extrapolation, we only include objects z$\textless\sim$1.2 in this comparison. As expected, as the UV dominates the SEDs of SLSNe (particularly at early times), we observe a more diminished range of energies here. However, we still observe a broad range in radiated luminosities of $\sim$1 dex, which still implies some range in explosion properties. 

Both sets of pseudo-bolometric light curves also highlight the diversity in rise and decline time behaviours within the sample. Several objects at higher redshift ($z>0.4$) rise much more swiftly than the $\sim20--30$ day rest-frame time identified as a \lq typical characteristic\rq\ of other literature SLSNe. 

It is possible that this broad range of physical characteristics could be encapsulated within the framework of a magnetar injection model. This model is capable of producing a broad range of light curve forms, given its dependency upon multiple parameters (including the mass and opacity of the ejected material, which naturally alter the diffusion time of photons through the ejecta). We explore the magnetar model in more detail in the following section.

\subsection{The Magnetar Model}\label{sect:magnetar}

The spin-down of a magnetar has been popularly invoked as the underlying energy source of SLSNe within the literature \citep[e.g.][]{Kasen2010,Woosley2010,Dessart2012,Inserra2013,Nicholl2013,Nicholl2017B}, which given its inherent flexibility, is capable of fitting a wide range of SLSN light curves. Indeed, it has previously been shown that the magnetar model provides a good fit to the main peaks of DES13S2cmm and DES14X3taz \citep{Papadopoulos2015;Smith2016}. Given the large diversity in light curve shapes present within the DES SLSNe, we next test the capabilities of the magnetar model against the whole sample.

We fit the magnetar model of \cite{Inserra2013} to our interpolated quasi-bolometric light curves, whose luminosity has the functional form (under the assumption of complete deposition of the magnetar energy into the expanding ejecta) of
\begin{multline}
L_{\mathrm{SN}}\left ( t \right )=e^{-(t/\tau_{\mathrm{m}})^{2}} \\ 2\int_{0}^{t/\tau_{\mathrm{m}}}
\left(\int 4.9\times10^{46} \left(\frac{B}{10^{14}~\mathrm{G}} \right)^{2}\left(\frac{P}{\mathrm{ms}} \right)^{-4} \frac{1}{(1 + t/\tau_{\mathrm{p}})^{2}}\right ) \\ e^{(t^{\prime}/\tau_{\mathrm{m}})^{2}}\frac{dt^{\prime}}{\tau_{\mathrm{m}}} \mathrm{~ erg~ s^{-1}},
\label{equation_magnetar}
\end{multline}
where $B$ and $P$ are the magnetic field strength and period of the magnetar respectively, $\tau_{\mathrm{p}}$ is the spin down timescale of the magnetar, and $\tau_{\mathrm{m}}$ is the diffusion timescale, which under the assumption of uniform ejecta density, can be expressed in terms of the mass, $M_{\mathrm{ej}}$, opacity, $\kappa$, and kinetic energy, $E_{\mathrm{k}}$ of the ejecta

\begin{equation*}
\tau_{\mathrm{m}} = 1.05(\beta c)^{-0.5}\kappa^{0.5}M_{\mathrm{ej}}^{0.75}E_{\mathrm{k}}^{-0.25} \mathrm{~ s},
\end{equation*}
where $\beta$ represents a normalisation constant, commonly taken to be 13.7 \citep{Arnett1982}. Following \cite{Inserra2013}, we assume an opacity of $\kappa = 0.1~\mathrm{cm}^{2} \mathrm{g}^{-1}$ (consistent with hydrogen-free ejecta). 

\begin{figure}
	\includegraphics[width=\columnwidth]{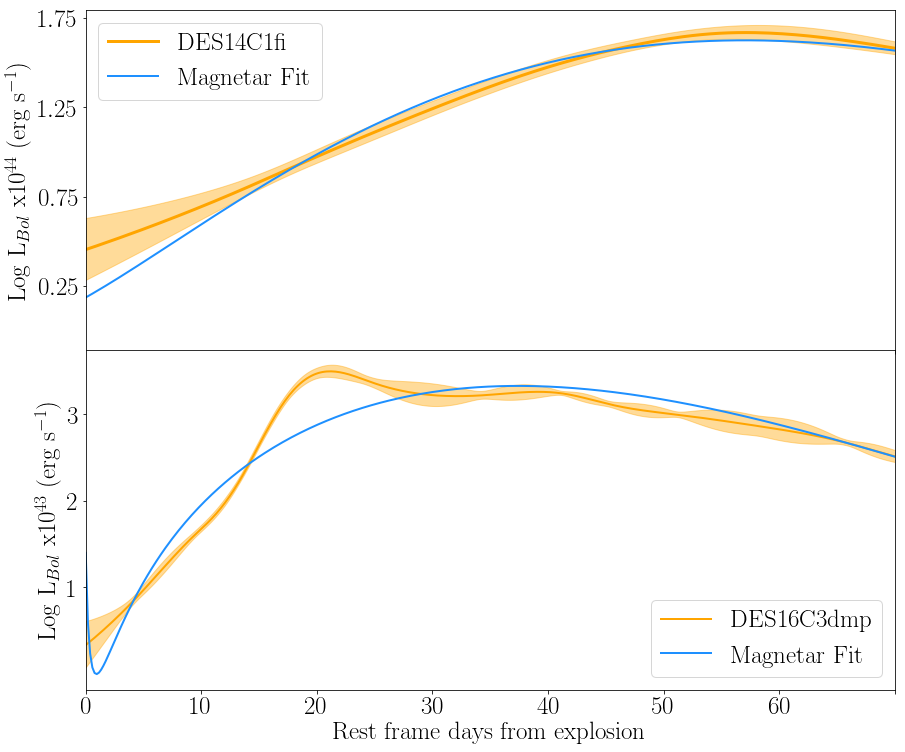}
    \caption{Example magnetar fits (blue lines) quasi-bolometric light curves of DES14C1fi and DES16C3dmp (orange lines). Both fits yield physical values to the magnetar fit parameters, and in some cases provide plausible replications of the bolometric light curves of the DES SLSNe (such as DES14C1fi). However others fair less well upon closer inspection, where the magnetar model is unable to encapsulate the unsmooth evolution of the quasi-bolometric light curve (e.g. DES16C3dmp).}
    \label{fig:magnetar}
\end{figure}

Due to temporal edge effects, some of the DES objects lack the light curve coverage required for a fair assessment of the magnetar model against their bolometric light curves. We therefore introduce a small time step as an additional fit parameter in \autoref{equation_magnetar}, in which we allow the start date of the SN to vary between $-10$ and $+10$ days to the start of the bolometric light curve. The best fitting magnetar properties are given for each SLSN with Table~\ref{tab:slsnproperties}. Whilst this model is able to fit a large fraction of the DES SLSN light curves, visual inspection of many of the fits reveals cases in which the smooth evolution of the magnetar model alone is unable to fully account for the wiggly evolution of the bolometric light curves (for e.g., see Figure~\ref{fig:magnetar}). Such cases could be a result of artefacts introduced from the GP interpolation, or they could be the result of multiple energy sources powering the light curve (e.g., magnetar + CSM interaction). We discuss this further in Section \ref{sect:dis}.

\begin{table*}
	\caption{The derived physical properties from the modified black body and magnetar models applied to the light curves of DES SLSNe.}
	\centering
	\begin{tabular}{l l l l l l} 
		\hline
		DES ID 	&  Peak T$_{\mathrm{BB}}$  	&Peak $R_{\mathrm{BB}}$ 	& $P_{\mathrm{spin}}$	 &$B$& $M_{\mathrm{ejecta}}$	 \\
		  		&   ($K$)  				& ($cm$) 						& ($ms$) 				& ($10^{14}G$)& ($M_{\odot}$)			 \\
		\hline
		DES13S2cmm	& 6.21e+03	& 7.99e+15  &         ...	&   ...	    &  ...  \\
		DES14C1fi	& 1.34e+04	& 7.99e+15 &         3.64	&3.12	 &17.32   \\
		DES14C1rhg	& 1.02e+04	&2.34e+15   &       4.95	&6.03	 &  3.19  \\
		DES14E2slp	& 8.52e+03	& 5.75e+15 &        ...	&...	 &...   \\
		DES14S2qri	& 1.054e+04	& 7.99e+15 &        4.95	&6.03	 &3.19   \\
		DES14X2byo	& 1.02e+04	& 7.99e+15  &       5.88	&4.99 & 3.87  \\
		DES14X3taz	& 1.06e+04	&  7.99e+15 &       5.30	&2.94	 & 19.66  \\
		DES15C3hav	& 9.30e+03  	& 2.78e+15 &    ...	&   ...	 &  ...  \\
		DES15E2mlf	& 1.81e+04	& 8.00e+15   &      5.60	&   2.68	 &  1.10  \\
		DES15S2nr	& 8.85e+03	&3.57e+15   &       10.52	&12.47	 & 6.17  \\
		DES15X1noe	& 1.13e+04	& 7.99e+15 &        7.45	&2.36	 & 1.15  \\
		DES15X3hm	& 1.25e+04 	&7.99e+15  &        5.63	&3.49	 & 4.18  \\
		DES16C2aix	& 8.25e+03	&7.99e+15  &         ...	&...	 &  ...  \\
		DES16C2nm	& 4.03e+03 	&6.99e+16  &        ...	&...	 & ...  \\
		DES15S1nog	& 1.36e+04	&2.31e+15  &        ...	&...	 &  ...  \\
		DES16C3cv	& 4.53e+03	&1.22e+16  &        ...	&...	 &  ... \\
		DES16C3dmp	& 1.24e+04	& 3.36e+15  &       18.27	&	9.98 &0.23   \\
		DES16C3ggu	& 8.82e+03	&8.92e+15   &       4.80	&1.05  & 16.24  \\
        DES17E1fgl	&7.38e+03  	& 7.34e+15  &       9.35	&3.60	 & 16.99  \\
        DES17X1amf	&	6.75e+03 & 1.61e+16 &       5.56	&2.22	 & 11.39  \\
        DES17C3gyp	&	1.07e+04 & 6.45e+15  &      6.29	&	2.94 & 11.57  \\
        DES17X1blv	& 5.87e+03	&  8.34e+15  &      ...	&	...  & ...  \\
		\hline
		\label{tab:slsnproperties}
	\end{tabular}
 \end{table*}

\section{Pre-peak Bumps}\label{bumpswiggles}
\label{sect:bumps_res}

There now exists strong evidence within the literature that the light curves of some SLSNe are multipeaked. Some events have shown signatures of re-brightening at later times, on both significant scales \citep[e.g. iPTF13ehe][]{Yan2015}, or on much more subtle small scales, manifesting as fluctuations within the decline of the main peak \citep{Nicholl2016B,Inserra2017}. A large fraction of SLSNe have exhibited bumps prior to the main peak of the light curve \citep{Leloudas2012,Nicholl2015Aa,Smith2016,Anderson2018}. Such features are not highly common to the bulk of the SLSN population, but as they may easily fall below the detection limits of shallower surveys, it is unclear whether these signatures are present in {\textit{all}} SLSNe. Understanding the nature of these features may provide the key to understanding the pre-explosion configurations of SLSN progenitors. Here we focus upon the presence of precursory peaks within the DES SLSN light curves.

Whilst the presence of bumps before the main peak of some SLSN light curves are well documented within the literature \citep{Leloudas2015,Nicholl2015Aa}, the precise nature of these bumps remains uncertain. \cite{Smith2016} highlighted the pre-peak bump observed within the light curve of DES14X3taz, being detected simultaneously in the DES $g,r,i,z$ bands 20 days prior to the rise of the main peak. Modelling of this bump favoured scenarios involving the shock cooling of an extended CSM located at $\sim400R_{\odot}$ from the progenitor. To date, this remains the best physically constrained SLSN pre-max bump. 

The shock cooling models of \cite{Piro2015} are highly dependant upon three parameters; the mass and radius of the circumstellar envelope ($M_{\mathrm{e}}$ and $R_{\mathrm{e}}$), and the mass of the core prior to explosion  $M_{\mathrm{c}}$ \citep[c.f.][]{Arcavi2017};

\begin{align}
L(t)= & 8.27\times10^{42}\Big(\frac{\kappa}{0.1 \mathrm{~g~cm^{-2}}}\Big)^{-1}\Big(\frac{v}{10^{9}\mathrm{~cm~s^{-1}}}\Big)^{2} 				\Big(\frac{R}{10^{13}\mathrm{~cm}}\Big) \notag\\
	 & \times \Big(\frac{M_{\mathrm{c}}}{M_{\odot}}\Big)^{0.01} \times \exp\left[-4.135\times10^{-11} \right. \notag\\
     &\left.  \times t\bigg(t\Big(\frac{v}{10^{9}\mathrm{~cm~s^{-1}}}\Big)+2\times10^{4}\Big(\frac{R}{10^{13}\mathrm{~cm}}\Big)\bigg) \right. \notag\\
     &\left.  \times\Big(\frac{\kappa}{0.1 \mathrm{~g~cm^{-2}}}\Big)^{-1} \Big(\frac{M_{\mathrm{c}}}{M_{\odot}}\Big)^{0.01}\Big(\frac{M_{\mathrm{e}}}{0.01M_{\odot}}\Big)^{-1} \right] \mathrm{erg~s^{-1}.}
\end{align}
\noindent
where $\kappa$ is the opacity of the CSM and $v$ is the expansion velocity of the shock breakout.

The degenerate nature of some of these parameters makes disentangling them for individual events complex, assuming that they all result from the same physical mechanisms. However, the bumps identified so far within the literature (with observations over the entire duration of the bump) appear to be of similar longevity, lasting around $\sim$10-20 days in the rest frame \citep{Leloudas2012,Nicholl2015Aa,Smith2016}, although with a spread in peak luminosities of $\sim$0.5 mag. If all SLSN bumps are the result of shock cooling from an extended circumstellar envelope then a similarity in bump duration may perhaps be indicative of very similar diffusion timescales, such that the combination of envelope mass and radius results in photons escaping the surrounding envelope over approximately the same time for all SLSN bumps. 

However, the model of \cite{Piro2015} used to model the bump of DES14X3taz is best constrained using multiband photometric observations, which few SLSNe within the literature possess. Fortunately the multiband photometry of the DES SLSNe presented here therefore offers the opportunity to test for the presence of similar \lq DES14X3taz-like\rq~ bumps within the pre-peak data. 

Given the pliability of the \cite{Piro2015} model in its capacity to replicate bumps over a variety of peak luminosities and durations, we test for the presence of bumps under the assumption that all bumps can be described by the model of \cite{Piro2015} and possess similar properties to those which have been documented within the literature. 

\begin{figure}
	\includegraphics[width=\columnwidth]{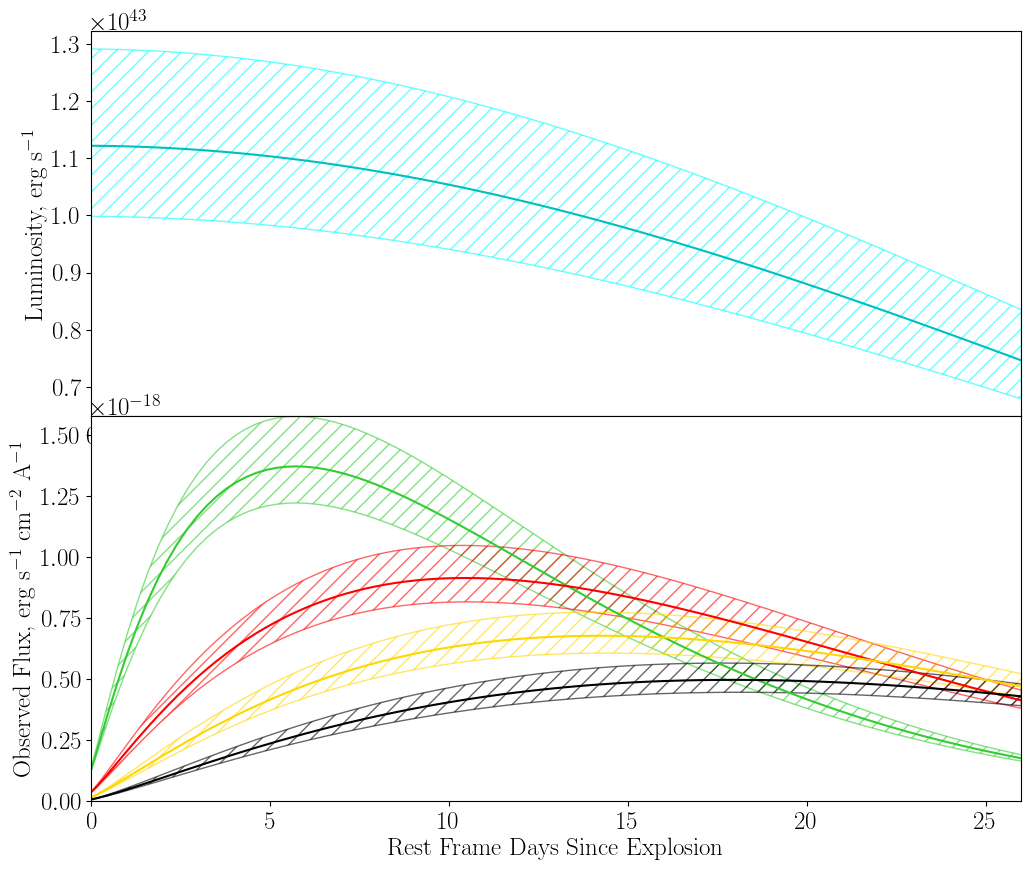}
    \caption{The shock outbreak into an extended CSM model of \citealt{Piro2015}, for which the mass and radius of the extended material have been fixed to the values found by \citealt{Smith2016}. We visualise the observed range of peak luminosities and corresponding observable fluxes in the DES filters for this physical scenario over a 5$\sigma$ range for a bump at the redshift of DES14X3taz ($z=0.608$). This basic model is scaled to the redshift of each SLSN in the DES sample used within an iterative Monte Carlo search of the pre-peak data for detections.}
    \label{fig:taz_bump}
\end{figure}

We therefore take the values of $M_{\mathrm{e}}$ and $R_{\mathrm{e}}$ as determined in \cite{Smith2016} as constant, such that we can place constraints upon the likely range of $M_{\mathrm{c}}$ that may be observed. $M_{\mathrm{c}}$ determines the efficiency of energy transfer from the explosion to the surrounding envelope, such that the change in SN energy with core mass becomes $\frac{E_{\mathrm{SN}}^{\prime}}{E_{\mathrm{SN}}} = \left(\frac{M_{\mathrm{c}}^{\prime}}{M_{\mathrm{c}}}\right)^{0.7}$, and thus is capable of changing the luminosity of the event. We find a core mass of 10.7$\pm$0.4$M_{\odot}$ for DES14X3taz.  In Fig.~\ref{fig:taz_bump} we show the corresponding range of peak bump luminosities (and observed DES $g,r,i,z$ fluxes) which may be generated using this scenario to match the observed spread in peak-bump $g$-band luminosities observed within the literature \citep[e.g. Figure 4 of][]{Smith2016}.

We then use this model to search for \lq X3taz-like\rq\ bumps within the DES SLSN data. To do this, we perform Monte Carlo simulations of SLSN bumps within the parameter space of DES14X3taz. Every realisation is then subtracted from the available pre-peak DECam photometry, moving the realised-model iteratively through the data out to $-60$ rest-frame days\footnote{This epoch represents the first detection of the earliest observed pre-peak bump within the literature, LSQ14bdq \citealt{Nicholl2015Aa}} from the main SN peak. The resulting residuals are then analysed for any $\geq$~3$\sigma$ detections in each band, with the requisite of that detections are found in a minimum of two or more bands at any epoch. For each event we are therefore able to determine a detection confidence from the ratio of detections to bump-realisations, which thus becomes the probability of having detected an \lq X3taz-like\rq\ for each transient. 

\begin{figure*}
	\includegraphics[scale=0.6]{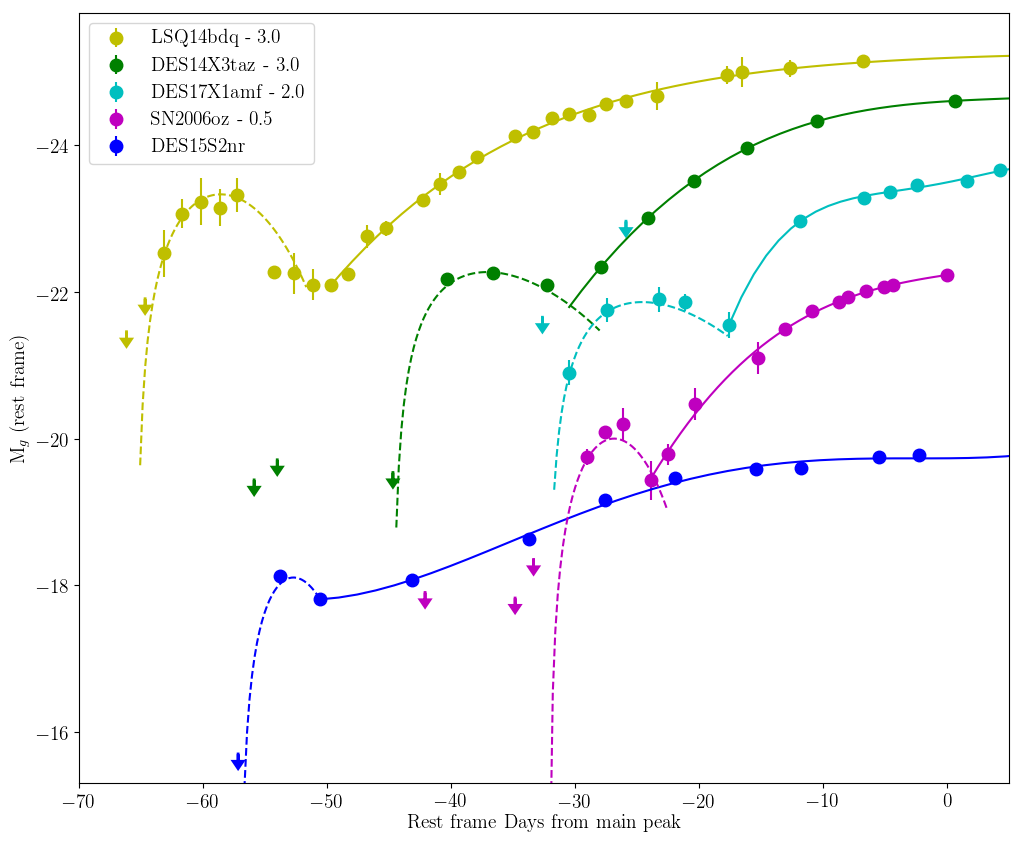}
    \caption{The rest-frame $g$-band lightcurves of DES SLSNe identified with pre-peak bumps (bar DES15C3hav), alongside the bumps of SN2006oz and LSQ14bdq \citep{Leloudas2012,Nicholl2015Aa}. Offsets in absolute magnitude have been applied for clarity and 3$\sigma$ limits are given where available. The bumps are fitted with the shock breakout within extended shell model of \citealt{Piro2015} (dashed lines), with the main SN peaks fitted with polynomials for clarity (solid lines). In the case of DES17X1amf, its high redshift ($z=0.92$) means the observed $z$-band bump displayed here is a better approximation of its near UV behaviour, rather than its behaviour at blue wavelengths.}
    \label{fig:g_band_bump}
\end{figure*}

\begin{figure*}
	\includegraphics[scale=0.68]{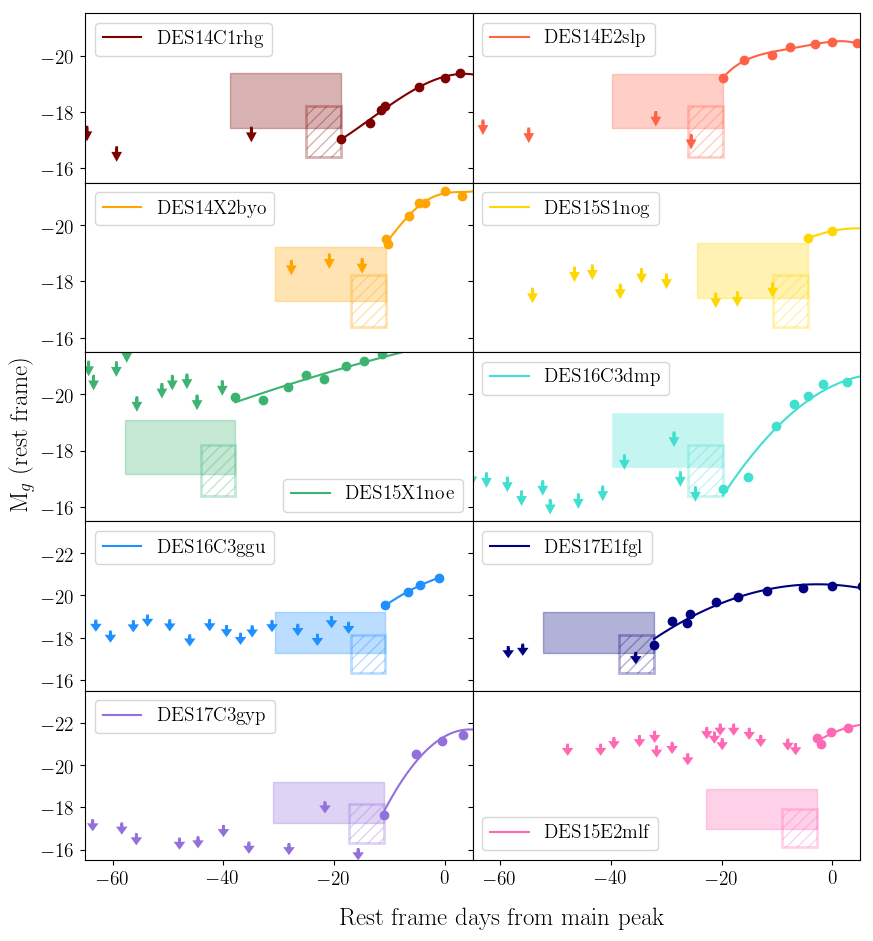}
    \caption{The rest-frame $g$-band light curves of DES SLSNe in which we do not identify a significant bump in the 60 days prior to the start of the main SN light curve.  We show the 3$\sigma$ limits for each SN and mark the parameter space in which we would have expected to observe the peak of both a \lq X3taz-like\rq\ and an \lq S2nr-like\rq\ bump in the rest-frame $g$-band (the filled and hatched regions respectively). With the exception of DES15X1noe and DES15E2mlf (whose high redshift precludes the identification of bumps of this magnitude range), we can rule out the presence of an \lq X3taz-like\rq\ bump down to limiting magnitudes of $M\sim-16$ for the more local SLSNe within this group ($z\sim~0.5$).}
    \label{fig:g_band_no_bump}
\end{figure*}

Of the DES SLSNe, 14 of the 22 events possess photometry prior to the main SN peak. Using this methodology, we identify significant bump-like signatures for 3 of these SNe, and firmly rule out the presence of bumps for 9/14 SLSNe with sufficient pre-peak data, with non-dectections down to limiting magnitudes of M$_{g}\sim$-16 for the more local SLSNe within this group ($z\leq0.5$). In the cases of DES15X1noe and DES15E2mlf, the identification of any bump-like features of DES14X3taz-like magnitude are precluded by the redshifts of these events ($z=1.18,1.86$ respectively). The peculiar case of DES15C3hav is discussed in Section \ref{sect:C3hav}, and we present the rest-frame $g$-band bumps in Figure \ref{fig:g_band_bump}, and the non-detections within Figure \ref{fig:g_band_no_bump}. 

Although identified as having detections consistent with the bump of DES14X3taz, from a wide range of bump durations introduced with the additional DES bumps (in particular, the rapid bump of DES15S2nr), it is clear that the bumps presented within Figure \ref{fig:g_band_bump} are not all best described by the parameters of DES14X3taz under the \cite{Piro2015} model. 

The identity of pre-peak bumps is still unclear. If still powered in a similar manner, the range in durations and luminosities could simply reflect a wider range in pre-explosion set-ups amongst SLSN progenitors, which may be captured within the flexibility of the shock outbreak model. However, other attempts to model bumps have provided equally plausible answers to the origin of SLSN bumps; for instance \cite{Margalit2018} have shown that within the constructs of models involving misalignment between a weak jet and the magnetic dipole of the magnetar powering it, the bump of LSQ14bdq can be explained by a mildly relativistic wind driven from the interface between a jet and the ejecta walls. The precise nature of SLSN bumps is unlikely to be solved without the addition of spectroscopic information. To date, only the potential candidate spectrum obtained during a bump epoch is that of SN2017egm \citep{Xiang2017,Nicholl2017}, where an early UV excess is detected within \textit{Gaia} data\footnote{Whether this detection occurs during the bump phase or during the very early stages of the main light curve is unclear \citep{Nicholl2017}.}, though falling below detection limits in the optical \citep{Bose2018}. The similarity of this spectrum to SLSN spectra near peak may cast doubt upon its identification as a true bump spectrum.

On the other hand, the confirmation of bump-less SLSNe within the DES data is also significant. A combination of poor cadence and shallow photometric limits have not conclusively ruled out the possibility that bumps are ubiquitous in the light curves of SLSN-I \citep{Nicholl2016A}. Here the deep, cadenced photometry of DES has provided the limits necessary to rule out the presence of \lq X3taz-like\rq\ bumps within the pre-peak data. 

\subsubsection{The unusual red \lq bump\rq of DES15C3hav}\label{sect:C3hav}

The Monte Carlo modelling to search for superluminous bumps also identified an early feature within the pre-peak data of DES15C3hav. Visual inspection of this feature proved it to be exceptionally red for an event at $z=0.376$ when compared to the colour of the main light curve and compared to other SLSN bumps. We observe clear detections within $i$ and $z$ bands, and some partial decline in the $r$ band at later times. 

\begin{figure}
	\includegraphics[width=\columnwidth]{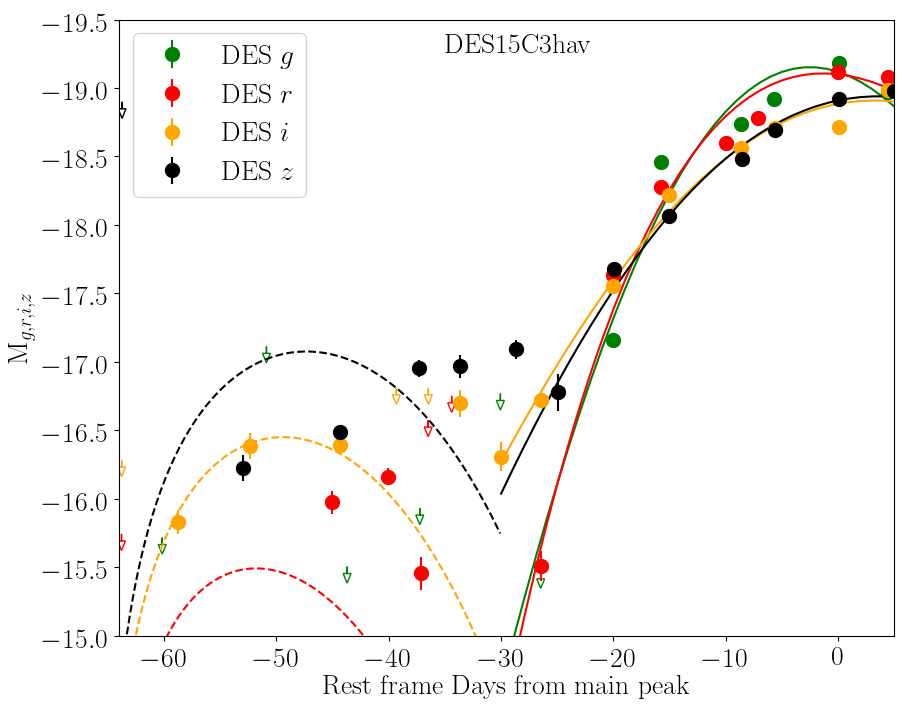}
    \caption{The unusual red bump of DES15C3hav identified at $\sim$50 days prior to the main peak of the transient as seen in the DES $g,r,i,z$ bands. The main peak of the SN in each band is fitted with a polynomial for clarity (solid lines), and 3$\sigma$ limits for each band are also shown. The dashed lines show the expected lightcurve of an \lq X3taz-like\rq~ bump with $\sim$3 magnitudes of additional reddening introduced. This arbitrary inclusion of additional extinction fails to fully capture  the behaviour of the bump of DES15C3hav, and requires more detailed modelling to describe its behaviour.}
    \label{fig:C3hav_bump}
\end{figure}

Although this red-feature, shown in Figure \ref{fig:C3hav_bump}, rises and declines on approximately the same timescale as a SLSN bump, it is $\sim2$ magnitudes fainter than the faintest bump identified within our sample, peaking at $M=-16.5$. When fit with a standard black body, it has an estimated black body temperature of $\sim$6000K. This red feature is strikingly different when compared to \lq typical\rq\ SLSN bumps (both here and within the literature), which are blue and hot \citep{Nicholl2015Aa,Smith2016}. It is even more unusual when the blueness of the main peak is also considered. 

It is possible that this feature is a normal SLSN bump, where the progenitor is surrounded by a shell of high opacity material which is subsequently destroyed during the shock breakout. Under this hypothesis, some properties of the DES15C3hav features could be explained, for instance, the late appearance of $r$-band flux could represent a change in opacity through which bluer light can begin to escape the surrounding shell. To test whether this feature could be described within the framework of the \cite{Piro2015} model, we overlay the bump of DES15C3hav with the \cite{Piro2015} fit to DES14X3taz. We attempt to account for the lack of bluer light during this bump phase by including A$_{V}$=3 magnitudes of additional extinction, passing this through the extinction laws of \cite{Cardelli1989} to determine the extinction in the DES bands, and plot the resulting bump within Figure \ref{fig:C3hav_bump}. Visually, this provides a poor match to the observed $r,i,z$-band behaviour of the bump of DES15C3hav, although it is sufficient to place the $g$-band light below the detection limits reached with DECam. However, its failure to encapsulate the observed lightcurve in other optical bands makes it unlikely this red feature is a product of shock breakout. 

The slow evolution and colour of this red feature are comparable to the \lq plateau\rq\ observed within the pre-maximum data of the more local SLSN, SN2018bsz \citep{Anderson2018}. The slow rising plateau of SN2018bsz, which endures for $>$26~days before the commencement of a more rapid rise to peak, was also found to be extremely red in colour, with a black body temperature of 6700K \citep{Anderson2018}. There are some similarities in the early time behaviour of this SN and DES15C3hav. Although the bump of DES15C3hav is redder (approximate restframe $u-g$ $>$0.59) and so fits a slightly cooler black body, the overall slow evolution and red colour mark these two events as distinct from other pre-peak bump events.

\section{Host Galaxies}
\label{sect:hosts}

The host galaxy environments of SLSNe have played an important role in understanding their progenitor origins. Several collective studies have shown that SLSN-I in particular exhibit a strong preference for faint host galaxies with low stellar masses and little star formation \citep{Neill2011,Lunnan2014,Lunnan2015,Angus2016,Perley2016} and generally sub-solar metallicities \citep{Perley2016,Schulze2016,Chen2017A}. These features common to the vast majority of SLSN host galaxies all heavily imply progenitors which are young and relatively massive (M$\geq$20M$_{\odot}$). 

If SLSNe are preferentially produced in low-metallicity environments, as we observe at low redshift, then one may expect to see an evolution of host galaxy properties with redshift, as at higher redshift, galaxies are typically less metal enriched for a given stellar mass \citep[as a result of a less chemically enriched early Universe, leading to an evolving mass-metallicity relationship,][]{Zahid2014,Ma2016}. The broad redshift range of the DES SLSNe sample allows us to test for the evolution of host properties out to $z\approx2$. 

We perform deep images stacks using images from the five-year DES-SN survey which have been selected such that they contain no SN light and exclude any taken under sub-optimal seeing and atmospheric conditions (Wiseman et al. \textit{in prep.}). This deep imaging allows us to detect the presence of host galaxy light down to limits of $m\sim25.5$--26.5 in the shallow and deep fields respectively. To avoid ambiguity in the case of multiple galaxies within the proximity of the SN, we consider the normalized elliptical radius of the galaxy in the direction of the SN (\lq directional light radius\rq, DLR \citealt[][]{Sullivan2006, Gupta2016}) of each galaxy. Hosts are identified through minimization of this DLR value. As per \cite{Gupta2016}, SNe are marked as \lq hostless\rq\ if the galaxies within the immediate environment have a DLR of $>4$, as this value minimises both the number of hostless events and the number of events with host confusion (Wiseman et al. \textit{in prep.}).

\begin{figure*}
	\centering
	\includegraphics[scale=0.55]{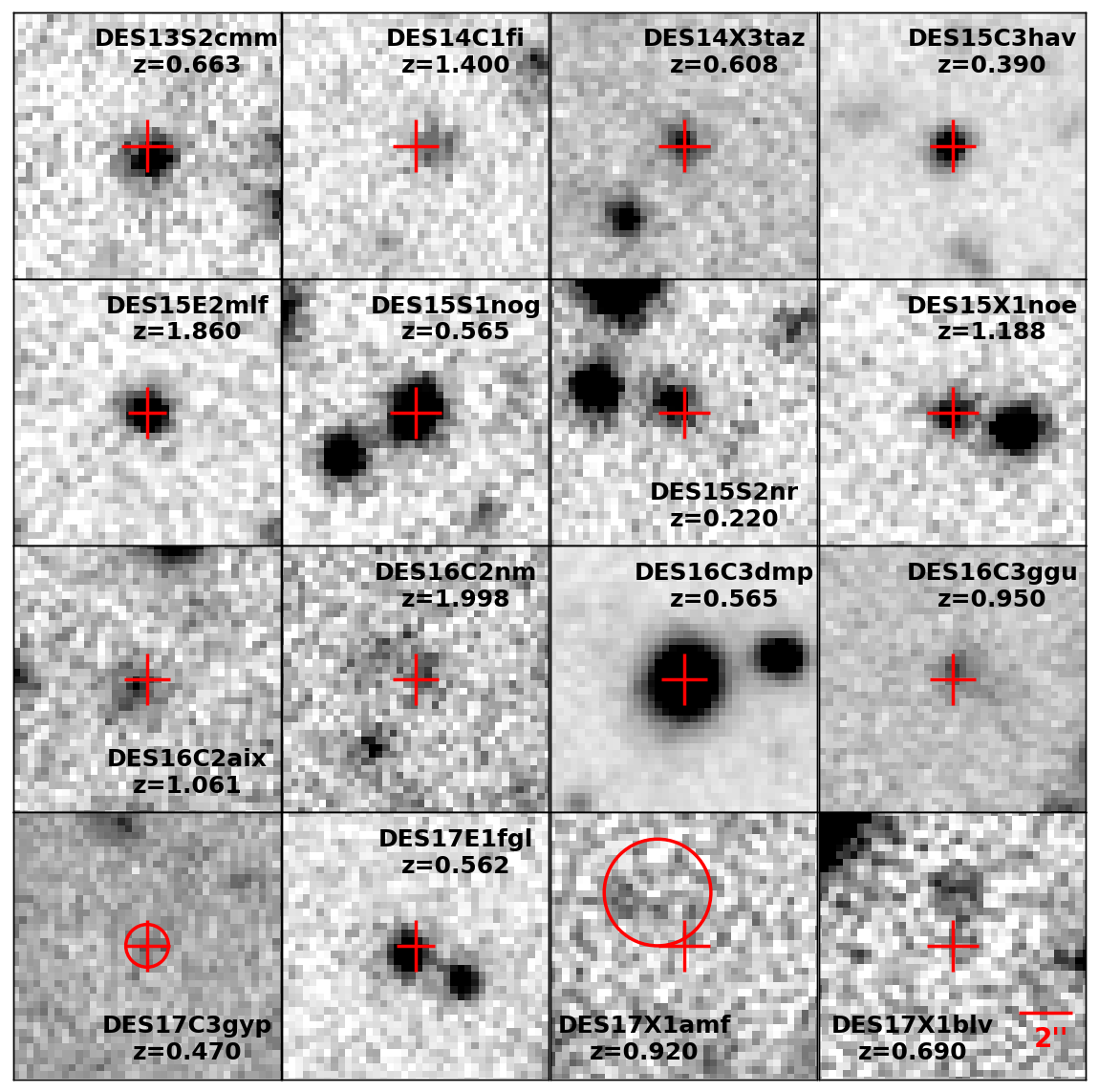}
    \caption{The stacked $r-$band imaging of the DES SLSN hosts. All images are 10\arcsec $\times$ 10\arcsec. Red crosses mark the location of the SN, and circles mark the loactions of fainter host galaxies where they fall within the limits of our imaging (but are still clearly detected within \textsc{sextractor}). We identify six events in the sample as being hostless down to the limits of our imaging; DES14C1rhg, DES14E2slp, DES14S2qri, DES14X2byo, DES15X3hm and DES16C3cv.}
    \label{fig:host_stamps}
\end{figure*}

We identify 16 host galaxies within the stacked images, and find no evidence of host galaxy light in $g,r,i,z$ for 6 of the SLSNe. The fraction of undetected hosts is similar to that seen within PanSTARRS \citep[without the aid of deep HST imaging,][]{Lunnan2013}, although considerably greater than that seen within shallower surveys such as PTF \citep{Perley2016}. Host galaxy stamps are presented within Fig.~\ref{fig:host_stamps} and their photometry is provided within Table ~\ref{tab:slsnhosts}.

\begin{table*}
	\caption{Photometry of the host galaxies of the DES SLSNe within their stacked $g,r,i,z$ band imaging. Where no host galaxy was detected, 3$\sigma$ limiting magnitudes are provided.}
	\centering
	\begin{tabular}{l l l l l} 
		\hline
		DES ID 	&  m$_{g}$  	& m$_{r}$ 	&  m$_{i}$ 	 & m$_{z}$ 	 \\
		\hline
        DES13S2cmm  & 23.92 $\pm$ 0.05 & 23.31$\pm$	0.04 & 22.98 $\pm$ 0.03 & 23.10	$\pm$ 0.05 \\
        DES14C1fi   & 24.52 $\pm$ 0.05 & 24.37$\pm$	0.06 & 24.21 $\pm$ 0.08 & 23.52	$\pm$ 0.05 \\
        DES14C1rhg  & $\textgreater$27.17 &	$\textgreater$26.63	& 	$\textgreater$ 25.37 & $\textgreater$25.66 \\
        DES14E2slp  & $\textgreater$26.99 &	$\textgreater$25.51	& 	$\textgreater$ 25.04 & $\textgreater$25.69 \\
        DES14S2qri  & $\textgreater$26.75 & $\textgreater$26.37	&   $\textgreater$ 25.97 & $\textgreater$26.26 \\
        DES14X2byo  & $\textgreater$25.93 &	$\textgreater$25.49	& 	$\textgreater$ 25.58 & $\textgreater$25.58 \\
        DES14X3taz  & 25.41 $\pm$ 0.09	&   24.69 $\pm$	0.04  & 24.39 $\pm$	0.05 & 24.40 $\pm$	0.05 \\
        DES15C3hav  & 24.60 $\pm$ 0.03	&   24.13 $\pm$	0.01  & 24.10 $\pm$	0.02 & 23.88 $\pm$	0.02 \\
        DES15E2mlf  & 23.35 $\pm$ 0.02	&   23.41 $\pm$	0.02  & 23.28 $\pm$	0.03 & 23.43 $\pm$	0.05 \\
        DES15S1nog  & 23.33 $\pm$ 0.03	&   22.54 $\pm$	0.01  & 22.16 $\pm$	0.02 & 22.06 $\pm$	0.02 \\
        DES15S2nr   & 23.64 $\pm$ 0.07	&   23.12 $\pm$	0.06  & 22.71 $\pm$	0.05 & 22.17 $\pm$	0.03 \\
        DES15X1noe  & 23.93 $\pm$ 0.04	&   23.80 $\pm$	0.04  & 23.41 $\pm$	0.04 & 23.44 $\pm$	0.05 \\
        DES15X3hm   & $\textgreater$26.67 & $\textgreater$26.28	&	$\textgreater$25.36 &	$\textgreater$25.63	\\
        DES16C2aix  & 24.69   $\pm$     0.07 &	24.49 $\pm$	0.07	& 24.37 $\pm$	0.09 &  24.11 $\pm$	0.09\\ 
        DES16C2nm   & 25.31   $\pm$	    0.10 &	25.08 $\pm$	0.12	& 24.94 $\pm$	0.13 &  25.45 $\pm$	0.25\\         
        DES16C3cv   &$\textgreater$ 26.46 &   $\textgreater$26.28	&	$\textgreater$25.94	 &	$\textgreater$25.87	\\
        DES16C3dmp  & 22.26    $\pm$	0.01 	& 21.55 $\pm$	0.01	& 21.29 $\pm$ 0.01	& 21.22 $\pm$	0.01    \\
        DES16C3ggu  & 25.21    $\pm$	0.06 	& 25.19 $\pm$	0.06	& 24.78 $\pm$ 0.05	& 24.49 $\pm$	0.05    \\
        DES17C3gyp  & 29.17    $\pm$	2.10   & 26.85 $\pm$	0.24	& 26.62 $\pm$ 0.12	& 26.29 $\pm$	0.11    \\
        DES17E1fgl  & 23.29    $\pm$	0.02 	& 23.06 $\pm$	0.03	& 22.64  $\pm$ 0.02	& 22.86 $\pm$	0.03    \\
        DES17X1amf  & 26.79    $\pm$    0.28 	& 27.09 $\pm$	0.43	& 26.44 $\pm$ 0.32	& 25.76 $\pm$	0.14    \\
        DES17X1blv  & 25.10    $\pm$	0.13 	& 25.42 $\pm$	0.20	& 24.40 $\pm$ 0.11	& 24.13 $\pm$ 0.10    \\
		\hline
		\label{tab:slsnhosts}
	\end{tabular}
 \end{table*}

We use \textsc{sextractor} \citep{Bertin1996} to determine the \textsc{mag\_auto} host galaxy magnitudes in $g,r,i~\& z$ bands. We compute the host galaxy stellar masses\footnote{Star formation rates of the DES SLSN host galaxies will be explored within later publications (D'Andrea et al. \textit{in prep.})} using the \textsc{z-peg} photometric-redshift software \citep{LeBorgne2002}, which uses the stellar population templates of \textsc{p\'egase.2} \citep{Fioc1997}. We assume assume a \cite{Kroupa2002} initial mass function and fix the redshift of each event as reported in Table \ref{tab:DESsample}.

\begin{figure}
	\includegraphics[width=\columnwidth]{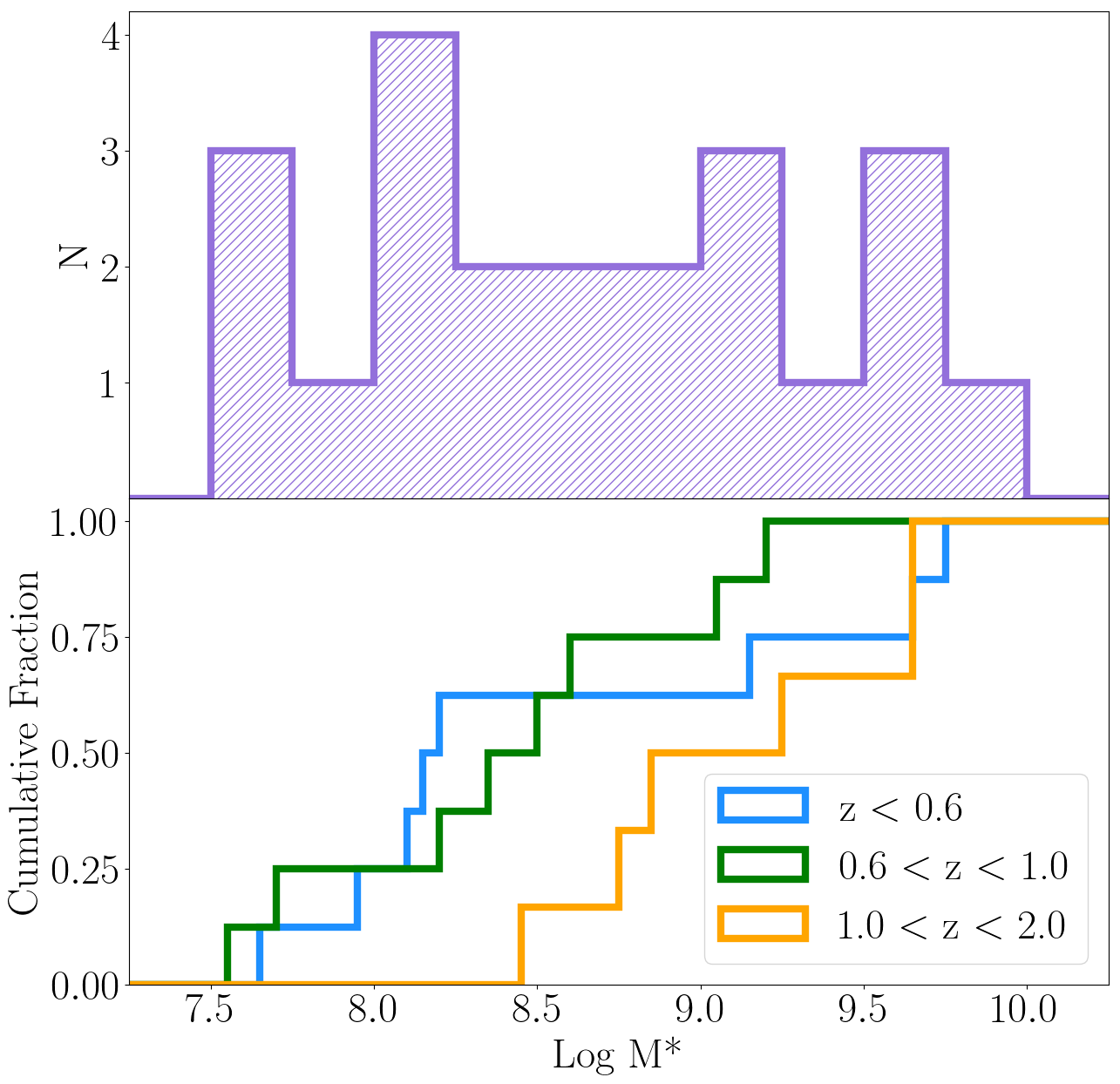}
    \caption{\textit{Top:} The host galaxy masses of the DES SLSN sample determined using \textsc{z-peg}. \textit{Bottom:} The mass function broken down into redshift bins of $z<$0.6, 0.6$< z <$1.0 and 1.0$< z <$2.0. There is arguably some evolution between the intermediate and high redshift bins, although small number statistics makes this tenuous.}
    \label{fig:host_mass}
\end{figure}

Fig.~\ref{fig:host_mass} shows the distribution of derived stellar masses, alongside the distribution when broken up into redshift bins of $z<0.6$, 0.6$< z <$1.0 and 1.0$< z <$2.0 (chosen such that we have approximately equal numbers of SNe per redshift bin). Overall, the stellar masses for the DES SLSNe are in accordance with those derived for other SLSN-I hosts, although we observe very little evolution of the mass function with redshift within our relatively small sample of hosts. Whilst visually there may be some slight evolution between intermediate to high-redshift, and is also statistically significant (a Kolmogorov-Smirnov test between the two samples yields a p-value of 0.02), this may be skewed by the small number of objects in each redshift bin. This potential evolution towards higher stellar masses at higher reshift would be consistent with the results of \cite{Schulze2016}, whose study of the properties of a larger sample of SLSN host galaxies find a redshift evolution of stellar mass consistent with the general cosmic evolution of starforming galaxies. Such behaviour is consistent with the evolution of the mass-metallicity relationship of field galaxies, supporting the notion of a metallicity bias in SLSN progenitor production.

However, the prerequisite for a faint host galaxy before targeting a candidate for spectroscopic followup may heavily bias our host galaxy results, particularly at low redshifts. We discuss this further within Section \ref{selecteffects}.

The properties of the host environment may also impact the properties of the resulting SNe. Such correlations are well established in other SNe classes -- for instance, SNe Ia in passive host galaxies are typically lower luminosity events with narrower, faster light curves \citep[e.g.,][]{Hamuy1996,Riess1999}. Other studies of SLSN environments have found tentative relations between the host galaxy enrichment and the derived properties of a magnetar spin-down fit to the bolometric light curve \citep{Chen2017A}.

\begin{figure}
	\includegraphics[width=\columnwidth]{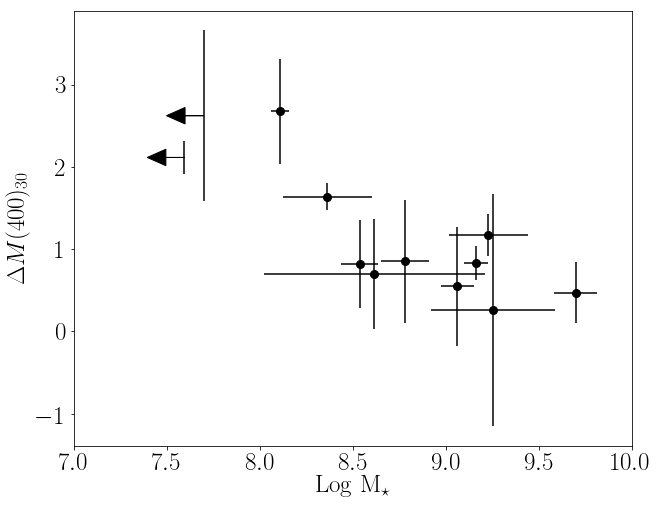}
    \caption{The apparent correlation between the evolution of DES SLSNe in the 4000\AA\ band between peak and 30 days and their host galaxy stellar masses. Although small numbers dominates our ability to draw firm conclusions, this relation could be indicative of stellar mass (and so metallicity) effects upon SLSN evolution.}
    \label{fig:host_correlation}
\end{figure}

We test for dependencies of the DES SLSN light curve properties with host galaxy mass. Whilst we find no significant correlation between host stellar mass and SN rise time, decline time, or colour, we do see a tentative negative correlation between stellar mass and the decline in the 4000\,\AA\ band between peak and 30 days, which we show in Figure \ref{fig:host_correlation}. This suggests that we observe more quickly evolving (i.e., redder at 30 days post-peak) events within lower mass host galaxies. Whilst we lack the sample size to draw more conclusions, this could be indicative of metallicity effects upon the progenitor stars; for instance, metal poor progenitors will have more optically thin ejecta, allowing energy to escape more easily, or perhaps within the paradigm of the relation between host metallicity and magnetar spin of \cite{Chen2017A}, a more rapidly spinning magnetar (for a given magnetic field strength) releases its energy much more quickly, resulting in a redder SN. 

\section{Discussion}\label{sect:dis}

\subsection{SLSN sample}

The definition of a SLSN has undergone many changes over the relatively short period since their discovery. The application of a luminosity threshold to select these events has for some time been found insufficient, occasionally leading to the inclusion of non-SLSN events such as tidal disruption events (TDEs, e.g. ASASSN-15lh \citealt{Leloudas2016}), and SN Ia-CSM events \citep[whose interaction bolsters both luminosity and duration of the transient,][]{Silverman2013}, resulting in an contaminated and poorly defined literature sample. Whilst homogeneously selected samples of SLSNe have begun to highlight a wider spread in their photometric properties \citep{Lunnan2017,DeCia2017}, they are still characterised as being generally more luminous than SNe Ia and CCSN events, and with slow light curve evolution.  

The sample of SLSNe within DES are all spectroscopically similar to other SLSNe within the literature; however, there exists a large amount of internal diversity within their light curves. Our distribution of peak luminosities spans almost four magnitudes, nearly 0.5 mag fainter than observed previously within the literature. \cite{DeCia2017} have shown that, based upon the sample of SLSNe, SNe Ic-BL and SNe Ibc observed within PTF, the volume-corrected distribution in peak luminosities for these transients shows no evidence that SLSNe are a separate population of events. The volume-weighted luminosity function presented here supports this, and broadens the range of peak luminosity space occupied by SLSNe. Such a continuous distribution could indicate similarities in the underlying explosion mechanisms, with some variation in progenitor set-up which could lead to the production of a brighter or fainter transient accordingly. 
 
Despite the somewhat limited light curve coverage resulting from the restricted length of the DES observing seasons, within the DES sample we see a wide range of rise times spanning a factor of $\sim4$, with some objects rising to peak bolometric luminosity in just 12 days in the rest frame (see DES15E2mlf, DES16C3dmp), whilst others taking nearly 50 days (DES15X1noe). A similarly broad spread is observed in decline timescales. 

A variety of general light curve properties were observed in both PTF and PANSTARRS samples. However, the well-constrained photometry of the DES survey reveals additional peculiarities, with many small-scale re-brightening events such as those identified by \citet{Inserra2017}, and the double-peaked re-brightening event of DES16C3cv, where the quasi-bolometric luminosity of the secondary peak is comparable to the brightness of the main peak. Whilst in isolated events, such deviations from a smooth light curve evolution may be less surprising, the fraction of SLSNe from DES which exhibit \lq atypical\rq\ behaviour is extremely high. It may be that this behaviour is common to all SLSNe, and has simply been highlighted by a combination of better photometric constraints and a higher redshift sample (where time-dilation allows these features to be more easily identified), as we begin to see deviations from \lq normal\rq\ behaviour in deeper photometric surveys \citep[cf.][]{Lunnan2017}.

The diversity in light curve properties has implications for the physical interpretation of SLSNe too. The application of a modified black body model has shown a peak temperature range between $\sim4000$ and $18000$ K, whilst in bolometric space this equates to a spread of $\sim1.5$\,dex in radiated energy. 
Magnetar energy injection has been popularly invoked to explain the observed luminosities and decline timescales of many SLSNe \citep[see][]{Kasen2010,Nicholl2013,Inserra2013,Nicholl2017, Dessart2018}. Whilst capable of replicating a wide range of transient lightcurves, this model presents a very smooth light curve profile, which by itself fails to capture the small scale variations seen within some light curves. We have shown that magnetars are incapable of describing some of the DES SLSNe, where the smooth evolution of the model cannot replicate the unusual light curves of some objects.  

There is a growing school of thought that magnetar injection cannot solely be responsible for the production of SLSNe, and may in fact be one of multiple energy production mechanisms that power the main light curve \citep{Wange2015}. For instance, the addition of CSM interaction would naturally explain bumpy lightcurves at late times, where the expanding ejecta reaches more distant shells of discarded stellar envelope as it moves outwards \citep{Inserra2017b}. The flexibility of the CSM model too, has appeal in describing why the features are not ubiquitous to {\textit{all}} SLSN light curves, as a low mass, low-opacity shell may not produce detectable wiggles within the late time light curve.

However, the delayed injection of energy from a central engine, rather than an instantaneous injection of energy at the point of collapse, as most models assume, may provide an explanation for the spread in peak luminosities observed for transients which are similar both spectroscopically and in their photometric evolution (i.e., Fig.~\ref{fig:DES_4OPS}). A time delay following the collapse allows some of the ejecta to fall back onto the newly formed compact remnant, reducing the ejecta mass and therefore the energy of the outward explosion \citep[see][]{Woosley2018}. This might explain the distribution of SLSNe within the 4OPS parameter space shown in Fig.~\ref{fig:DES_4OPS}, as the same physical mechanism causes the SN to evolve as they do, but is capable of producing a broad range in peak energies. 

\subsection{Bumps}

The enigmatic presence of pre-peak bumps within the light curves of some SLSNe-I has been recognised for some time. Their postulated origins vary wildly, from shock breakout into some form of extended stellar envelope or CSM \citep{Piro2015,Nicholl2015Aa,Smith2016,Vreeswijk2017}, to signatures of the shock created by the switch-on of a central magnetar at early times \citep{Kasen2016}, to emission from the collision of the SN ejecta with a close orbiting companion star \citep{Moriya2015}. Whatever their underlying physical nature, up unto this point it has been unclear whether these features are present within the light curves of {\textit{all}} SLSNe, where some fraction simply sit below the detection limits of their given surveys, or if they are simply not present at all. 

The depth of DECam imaging has allowed us to rule out the presence of \lq X3taz-like\rq\ bump-like features (down to the limits of their respective imaging) within the light curves of 10 SLSNe.  Furthermore, we have shown that a significant level of diversity (in both duration and luminosity) exists amongst the three bumps which are detected within the early time data. 

Under different physical setups, models involving the rapid heating and cooling caused by shock breakout within the progenitor envelopes prior to the main SNe are capable of encapsulating the wide variety of bump shapes observed within the DES sample. Degeneracies between the parameters within this model make it impossible to use any potential similarities in bump properties as hints of underlying progenitor properties, but with the inclusion of the DES sample, we clearly observe a much wider spread in bump properties than initially considered. 

Observations of precursory bumps within SN extends to include some much fainter core collapse SNe \citep[c.f.][]{Campana2006,Soderberg2008,Arcavi2011,Taddia2016,Barbarino2017,Taddia2018}, but why these events appear to manifest themselves most prominently within the light curves of more luminous supernovae is unclear. Unfortunately the DES dataset does not posses the cadence for more fine tuned models, typically sampling the light curve twice over the bump epoch. The addition of the DES SLSN bumps to the growing sample of observed bumps within the literature does not clarify the question of whether all bumps arise from the same underlying physical mechanism. However, this sample has significantly broadened the range of bump luminosities (which may, like the luminosities of the main peak, form a continuous distribution) and durations, which, if they are of one origin, suggests some flexibility in the allowed initial conditions prior to bump production. 

On the other hand, the lack of a pre-peak bump within some SLSN light curves is equally significant. Again, the source of energy production is common to all events, this could point to a variety of progenitor scenarios, in which sometimes the conditions are favourable for the production of a pre-peak bump, and in others less so. We note, that although our Monte-Carlo search revealed no bump prior to the start of the main peak, in some objects (particularly DES14X2byo and DES16C3dmp), we observe a small kink during the rise of the bluer bands in the light curve. It is possible that these small features may represent \lq belated pre-peak shock-cooling bumps\rq, where the extended material lies further from the progenitor than within a \lq classical\rq\ pre-peak bump (i.e. a \lq 14X3taz-like\rq) whose emission becomes merged with the main peak of the SN. Under a combined emission hypothesis, the appearance of prompt, belated or post-peak wiggles could simply represent CSM interaction across a range of radii from the progenitor.

The red pre-peak feature of DES15C3hav is also intriguing, in particular its contrast in colour and apparent temperature to the main peak of the SN. Our observations rule out the use of a shock breakout model to describe this feature. \cite{Anderson2018} suggest that the red plateau phase prior to the rise of the main peak of SN2018bsz could be explained by the late time injection of magnetar energy following a normal SN-Ibc event, which results in a more luminous SN at peak. This could be applicable to the behaviour of DES15C3hav, although it does not fully account for its strange colour evolution, in particular, the sudden decrease in $r$-band flux towards the end of the bump epoch.

\subsection{Selection Biases \& Future SLSN Searches}\label{selecteffects}

Given its cosmology-orientated science goals, the DES-SN survey is not an untargeted survey, with the spectroscopic follow-up primarily focused upon SN-Ia classification. The spectroscopic followup of our sample was prioritised based upon the \lq known\rq\ properties of SLSN from previous studies -- i.e., some combination of light curve and host galaxy properties, in particular for having long observed rise/decline times, and standing out as being several magnitudes brighter than any apparent host. Therefore our SLSN spectroscopic followup is likely to be incomplete. 

Under the assumption that these properties are typical of a SLSN, we can roughly estimate our spectroscopic completeness here based upon the highest redshift that we could have confirmed our faintest SLSN out to, which we find to be $z\approx0.63$. A comprehensive estimate of the number of spectroscopic completeness of DES SLSN followup will be presented within Thomas et al. (\textit{in prep.})

However, given the growing diversity in SLSN properties presented both here and within the literature \citep{Lunnan2017,DeCia2017}, this is likely to be an underestimate. The range of rest frame rise and decline times observed within the spectroscopically confirmed SLSNe in this work suggests that it is possible that our perquisite for a slow rise may result in some missing some fraction of low-$z$ events where the affects of time dilation are reduced, and therefore the SLSN evolves more quickly in the observer frame. However, without a spectroscopic redshift for every transient, the level of bias that this prerequisite introduces is difficult to quantify. 

The massive spiral galaxy host to SN2017egm \citep{Nicholl2017,Izzo2018,Chen2017B} highlights another bias in SLSN followup; although the fraction of SLSN within relatively massive host galaxies remains low, it is unclear whether this is the norm, and if SN2017egm-like events are simply missed by the majority of surveys due to the targeting of transients in fainter hosts. We test this faint-host bias within DES by searching for photometric candidate events which pass our other lightcurve criteria but are located within any host environment (bright or faint), and find that 66 unclassified transients pass this criteria. Given that we triggered spectroscopic follow up of 30 candidates under programmes designed to include SLSN events, this implies a followup completeness of \lq SLSN-like\rq\ events within DES of $\sim31$ per cent

Within this sample of 66 unclassified transients, we find that 12 meet our faint host criterion for spectroscopic followup. If this sample of candidate events were pure, this suggests that $\sim81$ per cent of possible SLSN events are missed due to their relatively \lq bright\rq\ host galaxies. Such a heavy bias in spectroscopic followup strategies could have implications for the progenitors SLSNe. Low luminosity, low mass host galaxies have previously been used as evidence for a young, massive population of progenitor stars \citep{Neill2011,Chen2013,Lunnan2014,Angus2016,Perley2016}. If SLSNe are indeed less localized to such exclusive environments, this widens the potential range of progenitor types significantly. 

However, this is highly likely to be an overestimate of the true fraction of missed SLSN events, as low-luminosity AGN outbursts may contaminate the sample, where any underlying variability from the source would likely fall within the noise of the survey, and thus lead to miss classification of the candidate event.

\section{Conclusions}\label{sect:con}
We have presented the light curves and classification spectra of 22 SLSNe from Y1-Y5 of DES. Objects in this sample were not initially selected based upon a luminosity cut, but rather based upon their slow light curve evolution, blue colour and faint host environment. They were classified as SLSNe based upon their spectroscopic similarity to other SLSNe identified within the literature, and this sample continues to add to the growing number of homogeneously selected SLSNe from wide-field surveys, although here we present the broadest redshift range yet within a spectroscopic sample, with a median redshift of 0.7 and reaching out to $z\sim2$. Analysis of the photometric properties of this sample show the following:

\begin{itemize}
\item The DES sample significantly extends the range of peak luminosities observed within the SLSN population, with the tail end of the distribution beginning to broach the peak luminosities of SN-Ia ($M\sim-19$).
\item We observe a broad spectrum of light curve characteristics (rise/decline times) and a large fraction of SLSNe which present \lq atypical\rq\ behaviour within the main peak of their light curves. This behaviour is difficult to encapsulate within the framework of a magnetar model alone. It is unclear whether this level of diversity is a direct result of multiple energy production mechanisms or a wide variety of progenitor properties.
\item We find signatures of pre-peak bumps are present within the light curves of three SLSNe within our sample. The range of bump durations and peak luminosities suggests again some variety in progenitor set-up prior to explosion.
\item We identify a particularly red feature prior to the main peak of DES15C3hav, which we are unable to describe under shock breakout models.
\item We can confirm the absence of pre-peak bumps within the first 60 days before the main peak within 10 SLSNe down to limits of $M\sim-16$. Although there are one or two cases in which we observe some excess of blue light (blue \lq kinks\rq) within the rise of the main peak, which may be signatures of a delayed pre-peak bump, the absence of bumps in a large fraction of light curves suggests they are no ubiquitous to all events of this class. 
\end{itemize}

It is clear that a magnitude limit cannot be applied to blindly select SLSNe within a survey. This work highlights the importance of multiband information in understanding their behaviour. In future surveys, an emphasis upon early-time spectroscopic followup may help to provide a better understanding of the mechanism(s) driving the main peak of the transient, but also of the nature of pre-peak bumps which are present within some SLSNe, and their connection to the main SN event. 

\section*{Acknowledgements}

CRA thanks Robert Quimby for providing the spectral templates used for classification purposes. 

We acknowledge support from EU/FP7-ERC grant 615929. CRA and MS thank the organisers and participants of the Munich Institute for Astro- and Particle Physics (MIAPP) workshop \lq Superluminous supernovae in the next decade\rq. 

Based on observations made with ESO Telescopes at the La Silla Paranal Observatory under programme IDs ESO 292.D-5013(A), 094.A-0310(B), 096.A-0536(A), 098.D-0057(A), 098.D-0057(B), 198.A-0915(A), 198.A-0915(C), 095.D-0797(A) and 097.A-0810(A).

Observations reported here were obtained at the MMT Observatory, a joint facility of the University of Arizona and the Smithsonian Institution. MMT observations were taken under programs 2014c-SAO-4 and 2015c-SAO-21.

Some of the observations reported in this paper were obtained with the Southern African Large Telescope (SALT). SALT observations were obtained under programs 2015-2-SCI-061, 2015-1-SCI-063 and 2017-2-SCI-049.

This paper makes use of observations taken using the Anglo-Australian Telescope under programs A/2013B/12 and NOAO 2013B-0317. 

This paper includes data gathered with the 6.5 meter Magellan Telescopes located at Las Campanas Observatory, Chile under programs CN2016B-16, and CN2017B-46. 

Some of the data presented herein were obtained at the W. M. Keck Observatory,  which is operated as a scientific partnership among the California Institute of Technology, the University of California and the National Aeronautics and Space Administration. The Observatory was made possible by the generous financial support of the W. M. Keck Foundation. Keck observations were taken under programs U021LA, U048LA and U150D.

Based on observations obtained at the Gemini Observatory, which is operated by the Association of Universities for Research in Astronomy, Inc., under a cooperative agreement with the NSF on behalf of the Gemini partnership: the National Science Foundation (United States), the National Research Council (Canada), CONICYT (Chile), Ministerio de Ciencia, Tecnolog\'{i}a e Innovaci\'{o}n Productiva (Argentina), and Minist\'{e}rio da Ci\^{e}ncia, Tecnologia e Inova\c{c}\~{a}o (Brazil). Gemini observations were obtained under program NOAO GS-2015B-Q-7. 

Based on observations made with the Gran Telescopio Canarias (GTC), instaled in the Spanish Observatorio del Roque de los Muchachos of the Instituto de Astrof{\'{i}}sica de Canarias, in the island of La Palma. GTC observations were obtained under programs 149-GTC70/14B and 163-GTC101/15B.

This research used resources of the National Energy Research Scientific Computing Center (NERSC), a U.S. Department of Energy Office of Science User Facility operated under Contract No. DE-AC02-05CH11231.

Funding for the DES Projects has been provided by the U.S. Department of Energy, the U.S. National Science Foundation, the Ministry of Science and Education of Spain, 
the Science and Technology Facilities Council of the United Kingdom, the Higher Education Funding Council for England, the National Center for Supercomputing 
Applications at the University of Illinois at Urbana-Champaign, the Kavli Institute of Cosmological Physics at the University of Chicago, 
the Center for Cosmology and Astro-Particle Physics at the Ohio State University,
the Mitchell Institute for Fundamental Physics and Astronomy at Texas A\&M University, Financiadora de Estudos e Projetos, 
Funda{\c c}{\~a}o Carlos Chagas Filho de Amparo {\`a} Pesquisa do Estado do Rio de Janeiro, Conselho Nacional de Desenvolvimento Cient{\'i}fico e Tecnol{\'o}gico and 
the Minist{\'e}rio da Ci{\^e}ncia, Tecnologia e Inova{\c c}{\~a}o, the Deutsche Forschungsgemeinschaft and the Collaborating Institutions in the Dark Energy Survey. 

The Collaborating Institutions are Argonne National Laboratory, the University of California at Santa Cruz, the University of Cambridge, Centro de Investigaciones Energ{\'e}ticas, 
Medioambientales y Tecnol{\'o}gicas-Madrid, the University of Chicago, University College London, the DES-Brazil Consortium, the University of Edinburgh, 
the Eidgen{\"o}ssische Technische Hochschule (ETH) Z{\"u}rich, 
Fermi National Accelerator Laboratory, the University of Illinois at Urbana-Champaign, the Institut de Ci{\`e}ncies de l'Espai (IEEC/CSIC), 
the Institut de F{\'i}sica d'Altes Energies, Lawrence Berkeley National Laboratory, the Ludwig-Maximilians Universit{\"a}t M{\"u}nchen and the associated Excellence Cluster Universe, 
the University of Michigan, the National Optical Astronomy Observatory, the University of Nottingham, The Ohio State University, the University of Pennsylvania, the University of Portsmouth, 
SLAC National Accelerator Laboratory, Stanford University, the University of Sussex, Texas A\&M University, and the OzDES Membership Consortium.

Based in part on observations at Cerro Tololo Inter-American Observatory, National Optical Astronomy Observatory, which is operated by the Association of 
Universities for Research in Astronomy (AURA) under a cooperative agreement with the National Science Foundation.

The DES data management system is supported by the National Science Foundation under Grant Numbers AST-1138766 and AST-1536171.
The DES participants from Spanish institutions are partially supported by MINECO under grants AYA2015-71825, ESP2015-66861, FPA2015-68048, SEV-2016-0588, SEV-2016-0597, and MDM-2015-0509, 
some of which include ERDF funds from the European Union. IFAE is partially funded by the CERCA program of the Generalitat de Catalunya.
Research leading to these results has received funding from the European Research
Council under the European Union's Seventh Framework Program (FP7/2007-2013) including ERC grant agreements 240672, 291329, and 306478.
We  acknowledge support from the Australian Research Council Centre of Excellence for All-sky Astrophysics (CAASTRO), through project number CE110001020, and the Brazilian Instituto Nacional de Ci\^encia
e Tecnologia (INCT) e-Universe (CNPq grant 465376/2014-2).




\bibliographystyle{mnras}
\bibliography{References} 




\appendix

\section{Appendix}
\FloatBarrier
\centering

\begin{table}
	\begin{tabular}{l l l  }

		\hline
		DES ID 	& Date Obs. & Telescope    \\
		 	&  &    (s) \\
		\hline
		DES13S2cmm	&	2013-11-21$\star$	&	VLT/Xshooter \\
		DES14C1fi	&	2014-09-21	&	AAT/AAOmega	\\
	            	&	2014-09-24	&	KECK/LRIS	\\
					&	2014-10-24	&	KECK/LRIS	\\		
					&	2014-10-30	&	AAT/AAOmega	\\		
					&	2014-10-30$\star$	&	VLT/Xshooter	\\		
		DES14C1rhg	&	2014-12-21	&	AAT/AAOmega   	\\		
					&	2014-12-24	&   AAT/AAOmega	    \\		
					&	2014-12-29$\star$	&	VLT/Xshooter   \\		
					&	2014-12-29	&	AAT/AAOmega    \\		
		DES14E2slp	&	2014-12-29$\star$	&  VLT/Xshooter		\\		
		DES14S2qri	&	2015-01-21$\star$	&	GTC/OSIRIS	\\	
        DES14X2byo	& 2014-10-18  & GTC/OSIRIS	    \\    
                    & 2014-10-23  & MMT/Blue Grating Spec.	    \\    
                    & 2014-10-24$\star$  & KECK/LRIS	    \\
                    & 2014-10-28  & AAT/AAOmega	    \\    
                    & 2014-11-16  & MAGELLAN  \\
                    & 2014-11-16  & MAGELLAN  \\
                    & 2014-11-20  & AAT/AAOmega	    \\    
                    & 2014-11-21  & AAT/AAOmega	    \\  
                    & 2015-01-21  & GTC/OSIRIS	    \\
		DES14X3taz	& 2015-01-26$\star$	&	GTC/OSIRIS	\\	
		            & 2015-02-06	&	GTC/OSIRIS	\\		
        DES15C3hav  & 2015-11-04 & VLT/Xshooter	\\
                    & 2015-11-09$\star$ & MMT/Blue Grating Spec.	\\
                    & 2015-11-12 & AAT/AAOmega	\\
                    & 2015-11-13 & AAT/AAOmega	\\
                    & 2015-12-12 & AAT/AAOmega	\\
        DES15E2mlf	& 2015-12-06 & GEMINI-S/GMOS	\\
                    & 2015-12-12 & AAT/AAOmega	\\
                    & 2015-12-14 & AAT/AAOmega	\\	
                    & 2015-12-15$\star$ & GEMINI-S/GMOS	\\
                    & 2016-01-14 & GEMINI-S/GMOS	\\	
		DES15S1nog & 2016-02-11$\star$ & GTC/OSIRIS \\ 	
        DES15S2nr	& 2015-09-19 &   VLT/Xshooter \\
                    & 2015-09-19 &   AAT/AAOmega \\
                    & 2015-09-21 &   AAT/AAOmega \\ 
                    & 2015-10-06 &   VLT/Xshooter \\
                     & 2015-10-10$\star$ &   KECK/LRIS	 \\
                     & 2015-10-19 &   SALT	\\
                     & 2015-10-20 &   SALT	\\
                     & 2015-11-03 &   SALT	\\
                     & 2015-11-04 &   VLT/Xshooter \\
                     & 2015-11-08 &   SALT	\\
                     & 2015-12-03 &   AAT/AAOmega \\
                    & 2015-12-11 &   KECK/LRIS	 \\
		DES15X1noe	&	2016-01-28$\star$	&	GTC/OSIRIS	\\
		DES15X3hm & 2015-09-09$\star$ & VLT/Xshooter  \\
		            & 2015-09-16 & AAT/AAOmega \\
                    & 2015-09-17 & AAT/AAOmega \\
                    & 2015-09-19 & AAT/AAOmega \\
       DES16C2aix  & 2016-10-10$\star$ & MAGELLAN	\\
                    & 2016-11-03 & AAT/AAOmega	 	\\
                    & 2016-11-26 & AAT/AAOmega	 	\\
                    & 2016-11-28 & AAT/AAOmega	 	\\
                    & 2016-11-29 & AAT/AAOmega	 	\\
       DES16C2nm   	& 2016-10-10$\star$ & 	MAGELLAN \\
                    & 2016-10-23 & 	VLT/Xshooter	     \\
                    & 2016-10-24 & 	KECK/LRIS	 \\
                    & 2016-10-24 & 	VLT/Xshooter	     \\
                    & 2016-11-21 & 	VLT/Xshooter	     \\
        \hline
        \label{tab:spec_obs}
		\end{tabular}
\end{table}
\FloatBarrier
\FloatBarrier

\begin{table}   
        \begin{tabular}{l l l } 
        \hline
		DES ID 	& Telescope & Date Obs.   \\
		 	&  &    (s) \\
		\hline
     DES16C3cv & 2016-09-25 & VLT/Xshooter	         \\
              & 2016-09-25 & AAT/AAOmega	         \\
              & 2016-10-10$\star$ & MAGELLAN	     \\
              & 2016-12-27 & KECK/LRIS	         \\
    DES16C3dmp & 2016-11-25 & AAT/AAOmega \\
                & 2016-11-28 & AAT/AAOmega \\
                & 2016-12-20$\star$ & VLT/Xshooter \\
    DES16C3ggu & 	2017-02-25$\star$	&	VLT/Xshooter \\
					
		DES17C3gyp 	& 2018-01-09 & SALT	\\
		& 2018-01-24$\star$ & AAT/AAOmega 	\\

        DES17E1fgl & 2017-12-22$\star$ & AAT/AAOmega \\
        DES17X1amf & 2017-09-22    & MAGELLAN	\\	
                    & 2017-10-22    & AAT/AAOmega	    \\	
                    & 2017-11-10$\star$    & MAGELLAN	\\	
                    & 2017-11-19    & AAT/AAOmega	    \\
                    & 2017-12-23    & AAT/AAOmega	    \\	
        DES17X1blv & 2017-11-10	  & MAGELLAN \\
                   &   2017-11-11$\star$ &  VLT/Xshooter	  \\
                   &   2017-11-19 &  AAT/AAOmega	  \\
        \hline
        \label{tab:spec_obs_ii}
	\end{tabular}
	\caption{A $\star$ indictaes the spectrum was used for classification.}
\end{table}
\FloatBarrier

\section{DES Photometry}\label{appendix_photom}


\begin{table}
\caption{DECam photometry of DES SLSNe. This is an abbreviated version of this table, shown here for content. The full table can be found in the
electronic version of the article.}
\begin{tabular}{cccccc}

\hline
     DES ID & MJD &  Phase & Band &    Mag &  MagErr \\
\hline
DES13S2cmm & 56534.28 & -16.91 &    g &  23.07 &    0.04 \\
DES13S2cmm & 56534.29 & -16.91 &    r &  22.79 &    0.04 \\
DES13S2cmm & 56534.29 & -16.90 &    i &  22.88 &    0.05 \\
DES13S2cmm & 56534.29 & -16.90 &    z &  22.69 &    0.05 \\
DES13S2cmm & 56538.32 & -14.48 &    g &  23.18 &    0.04 \\
DES13S2cmm & 56538.33 & -14.48 &    r &  22.60 &    0.03 \\
DES13S2cmm & 56538.33 & -14.47 &    i &  22.58 &    0.04 \\  
DES13S2cmm & 56538.33 & -14.47 &    z &  22.73 &    0.07 \\
DES13S2cmm & 56543.30 & -11.49 &    g &  22.92 &    0.03 \\
DES13S2cmm & 56543.30 & -11.49 &    r &  22.41 &    0.03 \\
DES13S2cmm & 56543.30 & -11.49 &    i &  22.56 &    0.04 \\
DES13S2cmm & 56543.30 & -11.48 &    z &  22.45 &    0.05 \\
 &  &  ...&      &     \\
 \hline
\end{tabular}
\end{table}

\bsp	
\label{lastpage}
\end{document}